\documentclass[aps,prd,showpacs,footinbib,twocolumn]{revtex4}
\usepackage{amsmath,amssymb,amsfonts}
\usepackage{graphicx,epsfig,graphics,rotate}
\usepackage{dcolumn}
\usepackage{bm}
\usepackage{dcolumn}
\usepackage{bm}
\DeclareGraphicsExtensions{.eps,.ps,.eps.gz,.ps.gz,.eps.Z,{}}

\newcommand{\be}{\begin{equation}}
\newcommand{\ee}{\end{equation}}
\newcommand{\gr}[1]{\mathbf{#1}}

\newcommand{\demi}{\frac{1}{2}}

\newcommand{\dbi}{\partial_{i}}
\newcommand{\dbj}{\partial_{j}}
\newcommand{\dhi}{\partial^{i}}
\newcommand{\dhj}{\partial^{j}}

\newcommand{\init}{\textrm{init}}
\newcommand{\vk}{\textbf{k}}
\newcommand{\vx}{\textbf{x}}
\newcommand{\vl}{\textbf{l}}
\newcommand{\mT}{{\cal T}}
\newcommand{\mTt}{{\cal T}_{\theta}}
\newcommand{\Dirac}{\delta_{\rm D}}

\def\ii{{\rm i}}
\def\dd{{\rm d}}
\def\mat{{\rm m}}
\def\baryon{{\rm b}}
\def\rad{{\rm r}}
\def\cdm{{\rm c}}
\def\HH{\mathcal{H}}
\def\EQ{\mathrm{eq}}
\def\LSS{\mathrm{LSS}}
\newcommand{\etas}{\eta_{\LSS}}
\newcommand{\sym}{{\hbox{sym.}}}
\newcommand{\fNL}{f_{_{\rm NL}}}
\newcommand{\fracb}{f_{_{\rm b}}}

\newif\ifjesuisfrancis\jesuisfrancistrue
\ifjesuisfrancis
\else
\fi

\begin{document}

\title{Cosmic microwave background bispectrum on small angular scales}

\author{Cyril Pitrou}
\email{pitrou@iap.fr}
 \affiliation{
              Institut d'Astrophysique de Paris, UMR7095 CNRS,
              Universit\'e Pierre~\&~Marie Curie - Paris,
              98 bis bd Arago, 75014 Paris, France,}
\affiliation{
              Institute of Theoretical Astrophysics,
              University of Oslo, 
              P.O. Box 1029 Blindern, 0315 Oslo, Norway,}
\author{Jean-Philippe Uzan}
 \email{uzan@iap.fr}
\affiliation{
              Institut d'Astrophysique de Paris, UMR7095 CNRS,
              Universit\'e Pierre~\&~Marie Curie - Paris,
              98 bis bd Arago, 75014 Paris, France,}

\author{Francis Bernardeau}
 \email{francis.bernardeau@cea.fr}
 \affiliation{CEA, IPhT, 91191 Gif-sur-Yvette c{\'e}dex, France,\\
              CNRS, URA-2306, 91191 Gif-sur-Yvette c{\'e}dex, France.}

\date{2 July 2008}
\begin{abstract}
This article investigates the non-linear evolution of cosmological
perturbations on sub-Hubble scales in order to evaluate the
unavoidable deviations from Gaussianity that arise from the
non-linear dynamics. It shows that the dominant contribution to
modes coupling in the cosmic microwave background temperature
anisotropies on small angular scales is driven by the sub-Hubble
non-linear evolution of the dark matter component. The
perturbation equations, involving in particular the first moments
of the Boltzmann equation for photons, are integrated up to
second order in perturbations. An analytical analysis of the
solutions gives a physical understanding of the result as well as
an estimation of its order of magnitude. This allows to quantify
the expected deviation from Gaussianity of the cosmic microwave
background temperature anisotropy and, in particular, to compute
its bispectrum on small angular scales. Restricting to equilateral
configurations, we show that the non-linear evolution accounts for
a contribution that would be equivalent to a constant primordial
non-Gaussianity of order $\fNL\sim25$ on scales ranging
approximately from $\ell\sim1000$ to $\ell\sim3000$.
\end{abstract}


 \maketitle

\section{Introduction}\label{sec1}

The cosmic microwave background (CMB) offers a unique window on
the physics of the early Universe, and in particular on
inflationary models. The angular power spectrum of the CMB
anisotropies has been extensively used to set constraints on the
shape of the inflationary potentials; see e.g.
Ref.~\cite{Komatsu:2008hk}. The statistical properties of the
temperature anisotropies and polarisation depend both on the
inflationary period during which they were created and on the
physics at play after Hubble-radius crossing and during the
recombination. At linear order in metric perturbations, those
latter physical processes amount to affect the metric
perturbations by a multiplicative transfer function. The
characteristic features observed in the temperature anisotropy
spectrum originate from the development of acoustic oscillations
that this transfer function encodes.  The overall amplitude of the
metric perturbation and its scale dependence are however
determined by the inflationary phase.

At linear order, the calculation of the transfer function - and
hence the detailed shape of the temperature power spectra - for
generic inflationary models requires the identification of the
relevant degrees of freedom during inflation (see e.g.
Refs.~\cite{1980PhRvD..22.1882B,1992PhR...215..203M,1996PThPh..95...71S}),
as well as a full resolution of the dynamics up to recombination
time. All these aspects are now fully understood (see e.g.
Refs.~\cite{PeterUzanBook,BernardeauBook} and references therein).

At this level of description, the metric perturbations are
linearised so that the non-linear couplings that are inherently
present in the Einstein equations are ignored. Therefore models
that predict Gaussian initial metric fluctuations  are expected to
induce cosmic fields with Gaussian statistical properties. This is
a priori the case for generic models of inflation. It has to be
contrasted to models with active topological defects - such as
cosmic strings - that have soon been recognised as a source of
large
non-Gaussianities~\cite{1994PhRvD..49..692P,Gangui:1996cg,Durrer:2000gi,Perivolaropoulos:1992gy}.
The current data however clearly favor only mild non-Gaussianities
although those might be larger than those induced by pure gravity
couplings. This is not the case for single field slow-roll
inflation for which it has been unambiguously shown in
Ref.~\cite{2003JHEP...05..013M} that it can produce only very weak
non-Gaussian signals, that are bound to be overridden by the
gravity induced couplings. It has however been realised that some
models of inflation might produce significant deviation from
Gaussianity in the context of multiple-field
inflation~\cite{2002astro.ph..6039K, 2002PhRvD..65j3505B,
2002PhRvD..66j3506B, 2003PhRvD..67l1301B, 2002PhLB..524....5L,
2004PhR...402..103B,
2004JHEP...04..006B,Bernardeau:2006tf,2004PhRvD..70d3533B} or with
non-standard kinetic terms~\cite{2004PhRvD..70l3505A}. The
question of the observation of primordial non-Gaussianities is largely open.\\

In general however, primordial deviations from Gaussianity are in
competition with the couplings induced during the non-linear
evolution of the cosmic fields. It has triggered general studies
aiming at characterising the bispectrum to be expected in the
observation of the cosmic microwave background temperature
anisotropies and polarisations whether it arises from inflation or
from subsequent effects.

This task is multi-fold. It requires a proper identification of
the mode couplings (at the quantum level) during the inflationary
phase - so taking into account the usual gauge freedom - as well
as a second order treatment of the post-inflationary evolution.
While the former has been set on firm
ground~\cite{2003JHEP...05..013M,2005PhRvD..72d3514W}, the latter
issue is still largely unexplored. This article proposes both
numerical and analytical insights into it.

Hereafter, we assume that on super-Hubble scales, the only
significant scalar perturbations are adiabatic and that they obey
a nearly Gaussian statistics~\footnote{Note that the choice of
$\zeta$ as the primordial field is not unique and one could have
chosen the Bardeen potential. With such a choice however the
$\fNL^{\zeta}$ incorporates only the inflation dependent couplings
- $\fNL^\zeta$ is proportional to the slow roll parameter in
single field inflation for instance. The other coupling terms
induce by the change of variable can be incorporated into
$\mT^{(2)}$; see Eq.~(\ref{fnltdef}) below.}. To be more precise,
they are described in Fourier space by a single variable
$\zeta_{0}(\vk)$, $\vk$ being a comoving wave-number, that
satisfies
\begin{equation}\label{zetaP}
 \langle \zeta_{0}(\vk_{1})\zeta_{0}(\vk_{2})\rangle
 =\Dirac(\vk_{1}+\vk_{2})\ P_{\zeta}(k_{1})
\end{equation}
and
\begin{widetext}
\begin{eqnarray}\label{zetaBispec}
\langle\zeta_{0}(\vk_{1})\zeta_{0}(\vk_{2})\zeta_{0}(\vk_{3})\rangle=
\hspace{0cm}2\Dirac(\vk_{1}+\vk_{2}+\vk_{3})\
\fNL^{\zeta}(\vk_{1},\vk_{2})
P_{\zeta}(k_{1})P_{\zeta}(k_{2})+\sym \ ,
\end{eqnarray}
where ``$\sym$'' stands for the two other terms obtained by
permutation of the wave-numbers. This defines the primordial power
spectrum $P_\zeta(k)$ and the primordial mode coupling
amplitude~\footnote{This is the expression for the bispectrum
obtained assuming $\zeta$ could be expanded as
$\zeta=\zeta_{G}+\fNL^\zeta\zeta_{G}\zeta_{G}$ where $\zeta_{G}$
is assumed to obey Gaussian statistics. This is not however a
valid description when the bispectrum originates from
multiple-field couplings or from quantum calculation. The formal
expression (\ref{zetaBispec}) is always valid though; see
Refs.~\cite{2004PhRvD..69f3520B,2005PhRvD..72d3514W}.}
$\fNL^{\zeta}$. Considering an observable quantity $\theta$
related to the perturbation variables, the effect of evolution can
generically be recapped~\footnote{Things are actually slightly
more complicated since usually observables cannot be decomposed
into 3D Fourier modes. The functions $\mTt^{(1)}(k)$ and
$\mTt^{(2)}(\vk_{1},\vk_{2})$ should then be thought as projection
operators. This is in particular the case for temperature
anisotropies and polarisations. This does not affect however the
general point we want to make in this introduction.} as
\begin{eqnarray} \label{FormalExp}
\theta(\vk)&=&\mTt^{(1)}(k)\, \zeta_{0}(\vk)
+\int\frac{\dd^3\vk_{1}\dd^3\vk_{2}}{(2\pi)^{3/2}}\,\Dirac(\vk\!-\!\vk_{1}\!-\!\vk_{2})
\mTt^{(2)}(\vk_{1},\vk_{2})\, \zeta_{0}(\vk_{1})
\zeta_{0}(\vk_{2}) +\dots\ ,
\end{eqnarray}
where $\mTt^{(1)}(k)$ is the linear transfer function and
$\mTt^{(2)}(\vk_{1},\vk_{2})$ is the second order transfer
function. $\theta$ can be thought as being e.g. the observed
temperature anisotropies, but it could also stand for the CMB
polarisation, or even cosmic shear surveys. When computing the
bispectrum of $\theta$ there will be a contribution from the mode
couplings induced by the second order transfer function
$\mT^{(2)}$ and the possible initial non-Gaussianities,
\begin{eqnarray}\label{FormalBispec}
\langle \theta(\vk_{1})\theta(\vk_{2})\theta(\vk_{3})\rangle&=&
2\Dirac(\vk_{1}+\vk_{2}+\vk_{3})\left[\fNL^{\zeta}(\vk_{1},\vk_{2})+
\fNL^{\theta}(\vk_{1},\vk_{2})\right]\mTt^{(1)}(k_{1})
\mTt^{(1)}(k_{2}) \mTt^{(1)}(k_{3})\
P_{\zeta}(k_{1})P_{\zeta}(k_{2})\nonumber\\
&&\qquad\qquad+\sym
\end{eqnarray}
\end{widetext}
where $\fNL^{\theta}$ is related to the second order transfer
function by
\begin{equation}\label{fnltdef}
\mTt^{(2)}(\vk_{1},\vk_{2})\equiv
\fNL^{\theta}(\vk_{1},\vk_{2})\,\mTt^{(1)}(|\vk_{1}+\vk_{2}|)\ .
\end{equation}
The full derivation of the details of $\mTt^{(2)}$ is a fantastic
task. It requires an understanding of the metric fluctuations
behaviour at second order, from radiation dominated super-Hubble
scales to matter dominated era at sub-Hubble scale, as well as a
comprehension of the physics of recombination - through the
Boltzmann equation - at a similar order. Such a task has been
undertaken by several authors~\footnote{Early derivations are to
be found in
Ref.~\cite{Maartens:1998xg,2006JCAP...06..024B,2007JCAP...01...19B}.
A more rigorous and comprehensive calculation -- including a
proper derivation of the Boltzmann coupling terms and taking into
account the polarisation effects -- is to be found in
Refs.~\cite{Pitrou:2007jy,pitrounew}.} and the multitude of
effects at play needs to be sorted out. So, the goal of this
article is not to provide an end to end calculation of
$\mTt^{(2)}$, but to show that on small scales one can extract the
dominant terms in order to get an insight into this physics at
second order.

Modes coupling due to gravitational clustering is, by far, not a
novel subject. It can be traced back to early works by Peebles
\cite{1980lssu.book.....P} where the function $\mT^{(2)}$ for the
non-linear sub-Hubble evolution of cold dark matter field (CDM)
during a matter dominated era was derived. General modes coupling
effects, within the same regime, has been extensively studied in
the eighties and nineties where a whole corpus of results has been
obtained (see e.g. Ref.~\cite{2002PhR...367....1B} for an
exhaustive review). On sub-Hubble scales, the second order mode
coupling function for the gravitational potential reads
\begin{eqnarray}\label{F2Peeb}
 && \fNL^{\Phi}(\vk_{1},\vk_{2})
      =\frac{k_{1}^2\,k_{2}^2}{\frac32\,H^2\,a^2\,\vert
        \vk_{1}+\vk_{2}\vert^2}\nonumber\\
 &&      \quad\times\left(\frac{5}{7}+\frac{1}{2}
       \frac{\vk_{1}\cdot\vk_{2}}{k_{1}^2}
      +\frac{1}{2}\frac{\vk_{1}\cdot\vk_{2}}{k_{2}^2}
      +\frac{2}{7}\frac{(\vk_{1}\cdot\vk_{2})^2}{k_{1}^2k_{2}^2}\right)
\end{eqnarray}
in the particular case of an Einstein-de Sitter universe (here $a$
is the scale factor and $H$ the Hubble parameter). This
well-established result proved to be useful for observational
cosmology. The angular modulation it exhibits has indeed been
observed in actual data sets; see e.g.
Ref.~\cite{2001ApJ...546..652S}.

The fact that on sub-Hubble scales, that is $k^2\gg H^2\,a^2$, the
non-Gaussianity is driven by the non-linearities of the CDM
sector, and that they can start developing even before equality is
one of the leading ideas of the present study. Indeed temperature
anisotropies on small angular scales -- i.e. beyond the first
acoustic peak -- mostly trace the gravitational
potential~\footnote{At least to some extent as we shall see in the
course of this paper.} long after it has entered a sub-Hubble
evolution and already during the matter dominated era. It is then
natural to expect that the temperature anisotropies should be
substantially determined by a form close to that of
Eq.~(\ref{F2Peeb}).

The goal of this paper is to evaluate how close we are from the
behaviour~(\ref{F2Peeb}) depending on scales, to which extent the
temperature anisotropies trace this form and finally to estimate
the amplitude of the temperature bispectrum on small angular
scales. In this work two approaches will be compared: a full
numerical integration of the second-order equations presented in
\S~\ref{sec2}, where the main approximation lies in the
modelisation of the Compton scattering collision term at second
order, see Eq.~(\ref{e.Hyp}), and an approximate analytical
resolution discussed in \S~\ref{sec3a}.

The bottom line of our analysis is that on small angular scales,
the density perturbation of the cold dark matter starts to
dominate the Poisson equation so at the time of decoupling we can
assume that the system is split in (1) the evolution of CDM
and (2) the evolution of the photons-baryons plasma which develops
acoustic oscillations in the gravitational potential determined by
the CDM component. As we shall demonstrate, at second order the dominant
term of the temperature fluctuations is driven by the second order
gravitational potential. Our approximation requires to consider a
regime in which the Silk damping is efficient, that is wave-modes
larger than the damping scales, hence corresponding to multipoles
roughly larger that 2000. This picture will be shown to be in
agreement with the numerical estimation (see \S~\ref{sec-num-an}).
We then proceed in \S~\ref{sec3} by a computation of the
bispectrum in which we show that, for equilateral configurations,
the non-linear dynamics has an amplitude equivalent to that of a
primordial non-Gaussianity with constant $\fNL$ of order 25. A
back-of-the-envelop argument allows us to understand the magnitude
of this number.

\section{Perturbation theory}\label{sec2}

This section is devoted to the presentation of the perturbation
equations, up to second order, and of the initial conditions used
in our study. We set the main notation and describe the
background dynamics in \S~\ref{sec2-1}, and we define the perturbation
variables in \S~\ref{sec2-2}. The perturbation equations and
initial conditions are then presented in \S~\ref{sec2-3}
and~\ref{sec2-4} respectively.

\subsection{The background dynamics}\label{sec2-1}

The background space-time is described by a
Friedmann-Lema\^{\i}tre metric with scale factor $a$ and cosmic
time $t$. It is convenient to rescale the scale factor such that
$$
 y\equiv \rho_\mat/\rho_\rad\ ,
$$
where $\rho_\mat$ and $\rho_\rad$ are the background matter and
radiation energy densities respectively. The matter energy density
can be decomposed as the sum of a cold dark matter component, that
does not interact with normal matter, and a baryonic component,
that can be coupled to radiation by Compton scattering prior to
decoupling. We thus set $\rho_\mat=\rho_\cdm+\rho_\baryon$ where
$\rho_\cdm$ and $\rho_\baryon$ refers to CDM and baryons
respectively. It follows that
$$
 \rho_\mat=\frac{\rho_\cdm}{1-\fracb}
$$
with $\fracb\equiv\Omega_{\baryon0}/\Omega_{\mat0}\approx 0.18$.
The Friedmann equation then takes the simple form
\begin{equation}
\label{e:hubble}
 \HH^2 = \HH_\EQ^2\frac{1+y}{2y^2}
\end{equation}
when we neglect the contributions of the spatial curvature and of
the cosmological constant, which are negligible for the whole
history of the Universe but very recently. $\HH\equiv a'/a$ is the
conformal Hubble parameter and a prime refers to a derivative with
respect to the conformal time $\eta$ defined by $\dd t =a\dd\eta$.
$\HH_\EQ$ is the value of the $\HH$ at equality, that is when
$y=1$.

The equation of state of the background fluid, composed of a
mixture of non-relativistic matter and radiation, is
$w=1/[3(1+y)]$ and the density parameters of matter and
radiation are
\begin{equation}
 \Omega_\mat=\frac{y}{1+y},\quad
 \Omega_\rad=\frac{1}{1+y}\
\end{equation}
and indeed $\Omega_\cdm=\Omega_\mat(1-\fracb)$.

Equality takes place at $y=1$, from which we deduce that
\begin{equation}
 y_0=1+z_\EQ = 3612\,\Theta_{2.7}^{-4}\left(\frac{\Omega_{\mat0}h^2}{0.15}\right),
\end{equation}
where $\Theta_{2.7}\equiv T_0/2.7$~K is the temperature of the CMB
today, and $h$ the value of the Hubble constant in units of
100~km/s/Mpc. Equation~(\ref{e:hubble}) evaluated today implies
that $\HH_\EQ\sim\HH_0\sqrt{2y_0}$ so that
\begin{equation}
 \HH_\EQ\sim0.072\ \Omega_{\mat0}h^2\,\mathrm{Mpc}^{-1}.
\end{equation}
The last scattering surface corresponds to a
redshift~\cite{Komatsu:2008hk}
\begin{equation}
 1+z_{\LSS} = 1090 \pm 1= y_0/y_{\LSS},
\end{equation}
and is mildly dependent of $\Omega_{\cdm0}$ and
$\Omega_{\baryon0}$. This implies that $y_\LSS\sim3.3$.

In Fourier space, a mode is super-Hubble when $k\eta\ll1$ and
sub-Hubble otherwise. The mode becoming sub-Hubble at equality
corresponds to a comoving wavelength of
\begin{equation}
 k_\EQ^{-1} = \HH_\EQ^{-1} = \frac{14}{\Omega_{\mat0}h^2}\,\mathrm{Mpc}
\end{equation}
if we choose units such that $a_0=1$.

We also introduce the parameter
\begin{equation}\label{e.defR}
 R=\frac{3}{4}\frac{\rho_\baryon}{\rho_\rad}
  =\frac{3}{4}\fracb\ y\ ,
 \end{equation}
which will be useful to describe the physics of the baryons-photons
plasma.

\subsection{Perturbation variables}\label{sec2-2}

We focus on the dynamics of scalar perturbations (see e.g.
Refs~\cite{Osano:2006ew,Lu:2007cj,Baumann:2007zm} for analysis of
vector and tensor modes generation at second order). In Newtonian
gauge, we can expand the metric as
\begin{equation}\label{metric}
 \dd s^2 = a^2(\eta)\left[-(1 + 2\Phi )\dd\eta^2
 + (1-2 \Psi)\delta_{ij}\dd x^{i}\dd x^{j}\right]\,,
\end{equation}
where $\Phi$ and $\Psi$ are the two Bardeen potentials.

The various fluids contained in the universe will be described at
the perturbation level by their density contrast $\delta$ and
their velocity field. For the latter, we decompose the time-like
tangent vector to the fluid worldlines according to
$$
 u^\mu=\frac{1}{a}(\delta^\mu_0+v^\mu)\ ,
$$
where the first term accounts for the background Hubble flow. The
perturbation $v^\mu$ is further decomposed as $v^\mu=(v^0,v^i)$
with $v^i=\partial^i v$ and $v^0$ is constrained by the
normalisation $u_{\mu}u^{\mu}=-1$.

When dealing with perturbations beyond first order, we assume that
the perturbation variables are expanded according to
\begin{equation}\label{decomposition-ordre2}
 X=X^{(1)}+\frac{1}{2}X^{(2)}\,,
\end{equation}
where $X^{(1)}$ satisfies the first order field equations while
the second order equations will involve purely second order terms,
e.g. $X^{(2)}$ as well as terms quadratic in the first-order
variables, e.g. $[X^{(1)}]^2$. Thus, there shall never be any
ambiguity about the order of perturbation variables involved as
long as we know the order of the equation considered, and
consequently we will usually omit the superscprit $(1)$ or $(2)$ which specifies the
order of the perturbation.
For a general discussions on second order perturbations and gauge
issues, we forward to Refs.~\cite{Pitrou:2007jy,Nakamura:2004rm,Bruni:1996im}

\subsection{Perturbation equations}\label{sec2-3}

In this article we shall focus on the CDM-radiation-baryons
system. Each component has a constant equation of state, that is
$P=w\rho$ with $w'=0$ so that $c_s^2=w$. Non-relativistic matter
is described by a pressureless fluid with $w_\mat= w_\baryon=
w_\cdm= 0$ and radiation satisfies $w_\rad=\frac13$.

In full generality, the evolution of each component can be
obtained from the Boltzmann equation satisfied by the distribution
function $f_a(x^\mu,p_\nu)$ for this matter component. The
stress-energy tensor can then be defined by integrating over
momentum as
$$
 T^{\mu\nu}_a(x^\alpha) = \int f_a(x^\alpha,p_\beta)\,  p^\mu p^\nu\pi_+(p)\ ,
$$
where $\pi_+(p)$ is the volume element on the tangent space in
$x^\alpha$ such that $p^\mu$ is non-spacelike and future directed
(see e.g. Refs~\cite{Uzan:1998mc,Pitrou:2007jy})

The first moment of the Boltzmann equation then gives a
conservation equation of the form~\cite{Uzan:1998mc}
\begin{equation}
\nabla_\mu T^{\mu\nu}_a = F^\nu_a\ ,
\end{equation}
where $F^\nu_a$ describes the force acting on the fluid labelled
by $a$ and satisfy $F^\nu_au_\nu=0$ and $\sum_aF^\nu_a=0$, which
is nothing but the action-reaction law (equivalently obtained from
the Bianchi identity). Projecting along and perpendicular to
$u^\mu$, we can extract respectively the continuity and Euler
equations.

\subsubsection{Linear order}

Linear order calculations are used in particular to set the source
terms of the second order equations. We closely follow the
standard calculations and the main ingredients are recalled here.
At linear order, the continuity equation for a fluid labelled by
$a$ takes the
form~\cite{PeterUzanBook,BernardeauBook,1992PhR...215..203M}
\begin{equation}\label{Eqcons1}
 \delta_a' +3 \HH (c_{s,a}^2-w_a)\delta_a+(1+w_a) \left(\Delta v_a -3 \Psi' \right) = 0\ ,
\end{equation}
while the Euler equation
\begin{equation}\label{Eqcons2}
 v_a'+ \HH (1-3 c_{s,a}^2) v_a+ \Phi + \frac{c_{s,a}^2}{1+w_a}\delta_a =
 \mathcal{F}_a -\frac{1}{6}\Delta\pi_a\ ,
\end{equation}
where $\pi_a$ is the contribution of the anisotropic pressure. In
deriving Eq.~(\ref{Eqcons2}), we have decomposed the force term as
$F_i=\partial_iF=\partial_i[(\rho+P)\mathcal{F}]$ and $F_0=0$.
Since $F^\mu_a$ vanishes at the background level, $\mathcal{F}_a$
is gauge invariant. $w_a$ and $c_{s,a}$ are respectively the
equation of state and sound speed of the component $a$.

In our analysis, we consider three components. Dark matter (label
$\cdm$) is described by a perfect fluid with $w_\cdm=0$ and
$\pi_\cdm=0$ interacting only through gravity
($\mathcal{F}_\cdm=0$). Baryons and photons are coupled through
Compton scattering so that $\mathcal{F}_\baryon$ and
$\mathcal{F}_\rad$ do not vanish. The action-reaction law (or
equivalently the conservation of the total stress-energy tensor of
matter) implies that $F_\rad=-F_\baryon$, from which we deduce
that
\begin{equation}\label{e.AR}
  \mathcal{F}_\rad = - R \mathcal{F}_\baryon\ .
\end{equation}
At linear order, it is easily shown that
\begin{equation}\label{e.FO1}
 \mathcal{F}_\rad =  \tau'(v_\baryon - v_\rad)\ ,
\end{equation}
where
\begin{equation}
 \tau'\equiv a n_e \sigma_T\ ,
\end{equation}
with $n_e$ being the free electrons number density and $\sigma_T$
the Thomson scattering cross-section. It follows that baryons will
be described by a fluid ($w_\baryon=0=\pi_\baryon=0$) interacting
with radiation. In general radiation enjoys a non-vanishing
anisotropic pressure ($\pi_\rad\not=0$) and should actually be
describe by the full Boltzmann hierarchy. For the linear order
calculations we choose to extend the fluid description by
including the eight first moments of this hierarchy, including
polarisation (see Appendix~\ref{AppRad} for these equations). Our
choices for the modelisation of the matter sector are summarised
in table~\ref{table1}.

\begin{table}
 \begin{tabular}{lcccc}
 \hline
 Component$\qquad$ & $w$ & $\mathcal{F}$ & $\pi$ & description \\
 \hline
 CDM ($\cdm$)  & 0 & 0 & 0 & fluid \\
 Baryons ($\baryon$)  & 0 & $\mathcal{F}_\baryon$& 0 & fluid \\
 Photons ($\rad$)  & $\,\,\frac13\,\,$ & $\,\,\mathcal{F}_\rad\,\,$ & $\,\,\pi_\rad\,\,$ &kinetic \\
          & &  & & (8 moments) \\
 \hline
 \end{tabular}
 \caption{Summary of the properties and descriptions of the matter components considered
 in our analysis.\label{table1}}
\end{table}

The Einstein equations reduce to the set
\begin{eqnarray}
  &&\Delta \Psi -3 \HH \Psi' -3 \HH^2 \Phi -\frac{3}{2} \HH^2 \sum_{a=\rad,\mat}
  \Omega_a \delta_a = 0\ ,\label{eq00}\\
&&\Psi'' + \HH^2 \Phi + \frac{1}{3}\Delta (\Phi-\Psi) + \HH \Phi' + 2 \HH \Psi' \nonumber\\
&&\quad \quad \quad +2 \HH'\Phi -  \demi \HH^2 \Omega_\rad \delta_\rad = 0\ ,\label{eqii}\\
 && \Psi - \Phi = \Omega_\rad\HH^2\pi_\rad\ ,\label{eqijst}\\
  && \Psi' + \HH \Phi + \frac{3}{2} \HH^2 \sum_{a=\rad,\mat} \Omega_a (1+w_a)v_a  = 0\label{eq0i}\ .
\end{eqnarray}
When the anisotropic pressure can be neglected, and in particular
in the tight coupling regime discussed below, Eq.~(\ref{eqijst})
implies that $\Phi=\Psi$. Note that in this analysis we actually
ignore the neutrinos the effect of which is thought to
be marginal on the qualitative results we will obtain.

\subsubsection{Second order}

At second order, any first order equation, schematically written
as $\mathcal{D}[X^{(1)}]=0$, of the first order perturbation
variables $X^{(1)}$ will take the general form
$$
 \mathcal{D}[X^{(2)}]=S
$$
where $S$ is a source term quadratic in the first order variables.

For the continuity and  Euler equations respectively read
\begin{widetext}
\begin{eqnarray}
 S_{c,a} &=& 2(1+w_a)\left\{ 6 \Psi \Psi' -
   \Phi\Delta v_a -\partial_i v_a\left[(1-3 c_{s,a}^2)\HH\partial^iv_a+2 \partial^i
      v_a'+2\partial^i\Phi-3\partial^i\Psi\right]\right\}\nonumber\\
      && +2(1+c_{s,a}^2)\left[3 \delta_a \Psi' -\partial_i(\partial^iv_a \delta_a)
\right]+\frac{(c_{s,a}^2)'}{1+w_a}\delta_a^2\ ,\label{e.Sc} \\
 \dbi S_{e,a} &=& -2 \frac{1+c_{s,a}^2}{1+w_a}\left[(\delta_a \dbi v_a)' + \HH (1 - 3
 w_a)\delta_a  \dbi v_a+\delta_a \dbi \Phi\right]
+ 2 \HH (1 - 3 c_{s,a}^2) (\Phi + 2 \Psi) \dbi v_a\nonumber\\
&& + 2 \Phi \dbi v_a' + 4 \Phi \dbi \Phi + 10 \Psi' \dbi v_a + 4
\Psi \dbi v_a'-2 \dbj \left(\dhj v_a \dbi v_a \right)
+2\frac{(c_{s,a}^2)'}{1+w_a}\left[\frac{\delta_a \dbi \delta_a}{3
\HH(1+w_a)}-\delta_a \dbi v_a\right]\ .\label{e.Se}
\end{eqnarray}
As long as CDM is concerned, these source terms and
Eqs.~(\ref{Eqcons1}-\ref{Eqcons2}) gives the full second order
evolution of the fluid. As already seen at first order, the fluid
equations for the baryons and photons must include interaction
terms, that is $\mathcal{F}^{(2)}_\rad$ and
$\mathcal{F}^{(2)}_\baryon$, that derive from the Compton
scattering collision term entering the Boltzmann equation for the
radiation.

In the baryon rest-frame, this collision term includes only two
types of
contributions~\cite{2006JCAP...06..024B,Dodelson:1993xz,pitrounew}.
First, there is a term involving first order perturbation
quantities and whose form is
\begin{equation}\label{e.Hyp1}
 \frac{1}{2}C^{(2)} \propto C^{(1)}\left(
 \frac{\delta n_e}{n_e} + \frac{\partial x_e}{\partial T}
 \frac{\delta T^{(1)}}{x_e}\right)\ ,
\end{equation}
where $C^{(1)}$ is the first order collision term. This
contribution involves the fluctuation of the visibility function,
that is of the electron density $n_e$ and of the ionisation
fraction $x_e$. It accounts for the fact that a hotter or denser
region decouples later. Its typical magnitude is of order
$4C^{(1)}\delta n_e/n_e$. Second, there is a term involving second
order perturbation variables and whose form is
\begin{equation}\label{e.Hyp2}
 \frac{1}{2}C^{(2)} \propto C^{(1)}\left[X^{(2)}\right]\ .
\end{equation}
The forces derived from these two terms will satisfy by
construction the action-reaction law~(\ref{e.AR}), and this holds
in any reference frame and at any order. This explains why the
computation is easily carried out in the baryons rest-frame~\cite{pitrounew}.

Then, when changing frame from the baryons rest-frame to the
cosmological frame, where the computations are actually carried
out, a second series of terms appears. They are of the form
$\tau'f^{(0)}\times[v^{(1)}]^2$ and $\tau'f^{(1)}\times[v^{(1)}]$,
where $f^{(0)}$ and $f^{(1)}$ are the background and first order
distribution functions as well as similar terms for the
polarisation (see Ref.~\cite{pitrounew} for the exact form of
these terms.).

Now, the contribution~(\ref{e.Hyp1}) is proportional to the
collision term at first order. This implies that it will thus be
negligible as long as tight coupling between baryons and photons is
maintained at {\it first order}, i.e. as long as $\tau'/k\gg1$. We
are thus left only with the contribution~(\ref{e.Hyp2}). This term
thus enforces the tight coupling regime at {\it second order}.
Then, as long as tight coupling is effective, it is obvious that
the second series of terms arising from the change of frames
should compensate each other to give a vanishing contribution. We
shall thus model the interaction term entering the Euler equations
by
\begin{equation}\label{e.Hyp}
 \mathcal{F}^{(2)} = \mathcal{F}^{(1)}[X^{(2)}]\ ,
\end{equation}
that is by assuming that it keeps the same functional form as at
first order. In conclusion, the continuity and Euler equations for baryons and photons with the interaction term~(\ref{e.Hyp})
are the exact fluid limit of the full Boltzmann equation at second
order as long as tight coupling is effective. This implies that at
second order, and similarly as at first order, the two tightly
coupled fluids are equivalent to a single perfect fluid; see
section~\ref{sec3a_heur}. Again, this stems from the
action-reaction law which implies that Eq.~(\ref{e.AR}) has to
hold at any order and in any reference frame and that there cannot
appear any external force acting on the resulting effective fluid
since it is only coupled to other matter components through
gravitation.

When $\tau'/k$ becomes of order unity, tight coupling stops being
effective. But thanks to Silk damping, the terms of the form
$\tau'f^{(0)}\times[v^{(1)}]^2$ and $\tau'f^{(1)}\times[v^{(1)}]$,
will still be negligible compared to the one we kept to obtain
Eq.~(\ref{e.Hyp}).\\

Let us now turn to the Einstein equations (\ref{eq00}-\ref{eq0i}).
Their source terms read respectively
\begin{eqnarray}
S_{1} &=& -8 \Psi \Delta \Psi -3 \dbi \Psi \dhi \Psi -3 \Psi'^{2}
       + 3 \HH^2 \sum_{a=\rad,\cdm,\baryon} \Omega_a (1+w_a)\dbi v_a \dhi v_a  -12 \HH^2
       \Psi^2\ ,\label{e.S1}\\
S_{2} &=&  4 \HH^2 \Psi^2 +\frac{7}{3} \dbi \Psi \dhi \Psi +
           \frac{8}{3} \Psi \Delta \Psi + 8 \HH \Psi \Psi'
    + 8 \HH' \Psi^2 + \Psi'^2 + \HH^2 \sum_{a=\rad,\cdm,\baryon} \Omega_a (1+w_a)\dbi v_a \dhi
    v_a\ ,    \\
S_{3} &=& -4 \Psi^2
     -\Delta^{-1}\left[2 \dbi \Psi \dhi \Psi + 3 \HH^2 \sum_{a=\rad,\cdm,\baryon}
      \Omega_a (1+w_a) \dbi v_a \dhi v_a \right]\nonumber \\
   && \qquad\qquad\qquad\qquad+ 3(\Delta \Delta)^{-1}\dbi \dbj\left[ 2 \dhi \Psi \dhj \Psi
    +  3 \HH^2 \sum_{a=\rad,\cdm,\baryon} \Omega_a (1+w_a) \dhi
        v_a \dhj v_a \right]\ ,\\
S_4 &=& 2 \HH \Psi^2 - 4 \Psi \Psi' + 2\dbi^{-1}(\Psi'\dbi \Psi)
    + 3 \HH^2 \sum_{a=\rad,\cdm,\baryon} \Omega_a \dbi^{-1}\left[(1+w_a)\Psi\dbi v_a
    -(1+c_{s,a}^2)\delta_a \dbi v_a \right]\ .\label{e.S4}
\end{eqnarray}
\end{widetext}
This provides all the source terms appearing at second order.

\subsubsection{Note on our conventions}

Since the first and second order equations are conveniently solved
in Fourier space, the quadratic terms in the source terms can be
written as a convolution on the wave-numbers $\gr{k}_1$ and
$\gr{k}_2$ such that $\gr{k}_1+\gr{k}_2= \gr{k}$. For simplicity
of notations we will write the integral factor of the convolution
as
$$
 \mathcal{C} \equiv \int \frac{\dd^3\gr{k}_1
 \dd^3\gr{k}_2}{(2 \pi)^{3/2}}\, \Dirac(\gr{k}_1+\gr{k}_2-\gr{k})\ .
$$
We also define $\mu=\gr{k}_1\cdot\gr{k}_2/(k_1 k_2)$. Unless
explicitly specified, we choose the convention of the Fourier
transform in which the factors of $(2\pi)^{n/2}$  are symmetric
for the Fourier transform and its inverse, $n$ being the dimension
of space.

\subsection{Initial conditions}\label{sec2-4}

To integrate this system of equations, we need to set the initial
conditions both for the first and second order variables deep in
the radiation era for super-Hubble modes at the initial time
$\eta_\init$, that is modes such that $k\eta_\init\ll1$.

\subsubsection{First order}\label{para.CI1}

At first order, we rely on the comoving curvature perturbation
which is constant on super-Hubble scales. It is
well-known~\cite{PeterUzanBook,BernardeauBook} that for a perfect
fluid with a time-dependent equation of state ($c_s^2 \neq w$), the comoving curvature perturbation, defined by
\begin{equation}
 \mathcal{R}^{(1)}=\Psi^{(1)} +\frac{2}{3(1+w)\HH}\left[\Psi'^{(1)} +  \HH\Phi^{(1)}\right]\ ,
\end{equation}
is conserved on super-Hubble scales for adiabatic perturbations.
Inflationary models predict the initial power spectrum of
$\mathcal{R}^{(1)}$ on those super-Hubble scales from which one
can deduce the power spectrum of the gravitational potential. If
we choose $\eta_\init$ such that the decaying mode is negligible
and $\Psi^{(1)}$ is constant on super-Hubble scales then, still
neglecting the anisotropic pressure,
\begin{equation}
 \mathcal{R}^{(1)}(\vk,\eta_\init)=\frac{5+3w}{3+3w}\Psi^{(1)}(\vk,\eta_\init)\ .
\end{equation}
Deep in the radiation era, this implies that
$$
 \Psi^{(1)}(\vk,\eta_\init)=\Phi^{(1)}(\vk,\eta_\init)=\frac{2}{3}
 \mathcal{R}^{(1)}(\vk,\eta_\init)\ ,
$$
for modes such that $k\eta_\init\ll 1$. Since the density contrast
of the total fluid
\begin{equation}
 \delta=\frac{1}{1+y}\delta_\rad+\frac{y}{1+y}\delta_\mat\ ,
\end{equation}
$\delta\simeq\delta_\rad$ deep in the radiation era. From
Eq.~(\ref{eq00}) we deduce that
$$
 \delta^{(1)}_\rad(\vk,\eta_\init) = -2\Phi(\vk,\eta_\init)\ .
$$
Now, assuming adiabatic initial perturbations, we must have
$$
 \delta^{(1)}_\cdm(\vk,\eta_\init) = \delta^{(1)}_\baryon(\vk,\eta_\init)
  = \frac{3}{4}\delta^{(1)}_\rad(\vk,\eta_\init)\ .
$$
Since baryons and photons are tightly coupled deep in the
radiation era, we deduce that
\begin{eqnarray}
 kv_\rad^{(1)}(\vk,\eta_\init) &=& kv_\baryon^{(1)}(\vk,\eta_\init) =
 kv_\cdm^{(1)}(\vk,\eta_\init) \nonumber \\
  &=& -\frac{1}{2}\Phi(\vk,\eta_\init)\ .\nonumber
\end{eqnarray}
This completely fixes the initial conditions for the set of first
order perturbation equations.

\subsubsection{Second order}\label{para.CI2}

At second order the previous procedure can be
generalised~\cite{2004CQGra..21L..65M,2005PhRvD..71f1301V}. It was
shown that on super-Hubble scales and for adiabatic perturbations,
the variable
\begin{widetext}
\begin{eqnarray}
\mathcal{R}^{(2)}&=&\Psi^{(2)} +\frac{2}{3(1+w)\HH}\left(\Psi'^{(2)} +
  \HH\Phi^{(2)} - 4 \HH \Psi^2 -  \frac{\Psi'^2}{\HH}\right)
+ \left(1 + 3 c_s^2\right)\left[\frac{\delta}{3(1+w)}\right]^2
+\frac{4}{3(1+w)}\delta \Psi\ ,
\label{Rsecondorder}
\end{eqnarray}
is a conserved quantity on super-Hubble scales for adiabatic
perturbations. Once the decaying modes are negligible so that
$\Phi^{(2)}$ and $\Psi^{(2)}$ are constant, we can express them in
terms of $\mathcal{R}^{(2)}$ as
\begin{eqnarray}
\Psi^{(2)} &=& \frac{1}{5+ 3 w} \Bigg\{ 3(1+w)\mathcal{R}^{(2)}
+4\frac{5+3w}{3(1+w)}\Psi^{(1)2} - 2
\Delta^{-1}\left[\frac{10+6w}{3(1+w)} \dbi \Psi^{(1)} \dhi
  \Psi^{(1)} \right]\nonumber\\
 &&\qquad\qquad  + 6 \Delta^{-2}\dhj \dbi
   \left[\frac{10+6w}{3(1+w)} \dbj \Psi^{(1)}\dhi \Psi^{(1)} \right]
   \Bigg\}\ ,\label{psiI}\\
\Phi^{(2)}&=&\Psi^{(2)} + 4 \Psi^{(1)2} +
\Delta^{-1}\left[\frac{10+6w}{3(1+w)} \dbi \Psi^{(1)} \dhi
\Psi^{(1)}\right]
 - 3 \Delta^{-2}\dhj \dbi \left[\frac{10+6w}{3(1+w)} \dbj
\Psi^{(1)} \dhi \Psi^{(1)}\right]\
.\label{Eq_lienPhiPsi_superHubble}
\end{eqnarray}

In the case where the initial perturbations have been generated
during a phase of one field inflation in slow-roll, it has been
shown~\cite{2003JHEP...05..013M} that, for the variable defined in
Eq.~(\ref{Rsecondorder}), $\mathcal{R}^{(2)}\simeq
-2\mathcal{R}^{(1)2} +\mathcal{O}(\textrm{slow-roll parameters})$.
Under this hypothesis, we can use the constancy of
$\mathcal{R}^{(2)}$ on super-Hubble scales and
Eqs.~(\ref{psiI}-\ref{Eq_lienPhiPsi_superHubble}) to derive the
initial conditions satisfied by the two Bardeen potentials at
second order at an initial time $\eta_\init$ deep in the radiation
era. Up to small corrections of the order of the slow-roll
parameters, they are given by
\begin{eqnarray}
\Psi^{(2)}(\eta_\init) &=&  -2\Psi^{(1)2}(\eta_\init) -
\Delta^{-1}\left[ \dbi \Psi^{(1)} \dhi \Psi^{(1)}
\right]_{\eta=\eta_\init} + 3\Delta^{-2}\dhj \dbi
\left[ \dbj \Psi^{(1)}\dhi \Psi^{(1)} \right]_{\eta=\eta_\init}\ ,\\
\Phi^{(2)}(\eta_\init) &=&  +2\Psi^{(1)2}(\eta_\init) +2
\Delta^{-1}\left[ \dbi
  \Psi^{(1)} \dhi \Psi^{(1)} \right]_{\eta=\eta_\init}
  - 6\Delta^{-2}\dhj \dbi \left[ \dbj \Psi^{(1)} \dhi
\Psi^{(1)} \right]_{\eta=\eta_\init}\ . \label{psiIr}
\end{eqnarray}
\end{widetext}

We also need to determine the initial conditions for the energy
density contrasts and the velocities of the different matter
components.  Again, we assume adiabatic initial conditions, which
means that the total fluid behaves like a single fluid. While at
linear order the pressure and density perturbations are simply
related by the sound speed
\begin{eqnarray}
\delta P^{(1)}&=&c_s^2 \delta \rho^{(1)}\ ,
\end{eqnarray}
at second order we have
\begin{eqnarray}
\delta P^{(2)}&=&c_s^2 \delta \rho^{(2)}+\frac{(c_s^2)'}{\bar
\rho'} (\delta \rho)^2\ .
\end{eqnarray}
This implies that the adiabaticity conditions at second order
reads
\begin{eqnarray}\label{CI2abiab}
\frac{\delta_\mat^{(2)}}{3}&=&\frac{\delta_\rad^{(2)}}{4}-\left(\frac{\delta_\rad^{(1)}}{4}
\right)^2=\frac{\delta_\rad^{(2)}}{4}-\left(\frac{\delta_\mat^{(1)}}{3}
\right)^2\ ,
\end{eqnarray}
where use of the first order adiabaticity condition was made to get
the last equality. It can also be shown from the perturbation
equations that the condition~(\ref{CI2abiab}) remains valid on
super-Hubble scales, hence giving a conservation of the
baryon-photon entropy on large scales, exactly as at linear order.

Following the same procedure as at first order, we first use the
Poisson equation~(\ref{eq00}) and Eq.~(\ref{eq0i}) at second order
to get that
$$
 \delta_\rad^{(2)}(\eta_\init) = -2\Phi^{(2)}(\eta_\init)
  + 8 \Phi^{(1)}(\eta_\init)\Phi^{(1)}(\eta_\init)\ ,
$$
and
\begin{eqnarray}
 kv_\rad^{(2)}(\vk,\eta_\init) &=&
 \frac{1}{2}k\eta_\init\left[
 -\Phi^{(2)}(\vk)
 \right.\nonumber\\
 &&\hspace{-2cm}+\left. \Phi^{(1)}(\vk_1)\Phi^{(1)}(\vk_2)
 \left(2 - 3\frac{\vk}{k}\cdot\left(\frac{\vk_1}{k_1}+\frac{\vk_2}{k_2}\right) \right)
 \right]_{\eta_\init}.\nonumber
\end{eqnarray}
Then using Eq.~(\ref{CI2abiab}) we conclude that
\begin{eqnarray}
 v^{(2)}_\cdm(\eta_\init) &=& v^{(2)}_\baryon(\eta_\init) = v^{(2)}_\rad(\eta_\init) \
 ,\nonumber\\
 \delta^{(2)}_\cdm(\eta_\init) &=&
 \delta^{(2)}_\baryon(\eta_\init)=
 \frac{3}{4}\left[\delta^{(2)}_\rad - \Phi^{(1)}\Phi^{(1)}
 \right]_{\eta_\init}\ .
\end{eqnarray}
This completely fixes the initial conditions for the set of second
order perturbation equations.

\subsection{Integrating the evolution equations}\label{Integrations}

Working in Fourier space, we first integrate the first order
equations, that is
\begin{itemize}
 \item Eqs.~(\ref{Eqcons1}-\ref{Eqcons2}) for CDM and baryons assuming
 a source term of the form~(\ref{e.FO1}) for baryons;
 \item the first eight moments of the Boltzmann equation for the
 radiation including the contribution of the polarisation. These
 equations are detailed in Appendix~\ref{AppRad};
 \item the Einstein equations~(\ref{eq00}-\ref{eq0i}).
\end{itemize}
The initial conditions are detailed in \S~\ref{para.CI1}.
Technically, we have recast all these equations in order to use
$y$ as time variable and we remind that the time of decoupling is
of the order of $y=3$. We also remind that the modes of interest,
that is $k>k_{\rm eq.}$, becomes sub-Hubble at $y<1$. This first
integration thus  allows us determine all the first order
perturbations as a function of time and wave-number. An example of
the results of the first order integration is presented on
Fig.\ref{fig3} (Appendix~\ref{appC}).

Now, at second order, we integrate the same system of equations
but supplemented by the source terms which are determined by the
solutions of the previous integration. We thus solve
\begin{itemize}
 \item Eqs.~(\ref{Eqcons1}-\ref{Eqcons2}) for CDM and baryon with
 the source terms~(\ref{e.Sc}) and~(\ref{e.Se}) respectively. We
 recall our main hypothesis which states that the coupling of baryons to
 radiation can be described by the interaction term~(\ref{e.Hyp});
 \item radiation is described by the first four moments of the
 Boltzmann equation with the hypothesis~(\ref{e.Hyp}) for the
 collision term;
 \item the Einstein equations~(\ref{eq00}-\ref{eq0i}) with the source
 terms~(\ref{e.S1}-\ref{e.S4}).
\end{itemize}
The initial conditions are detailed in \S~\ref{para.CI2} so that
we are finally able to compute the evolution of the perturbation
variables with $y$ for any wave-number.

\begin{figure}[htb]
\center
\includegraphics[width=8cm]{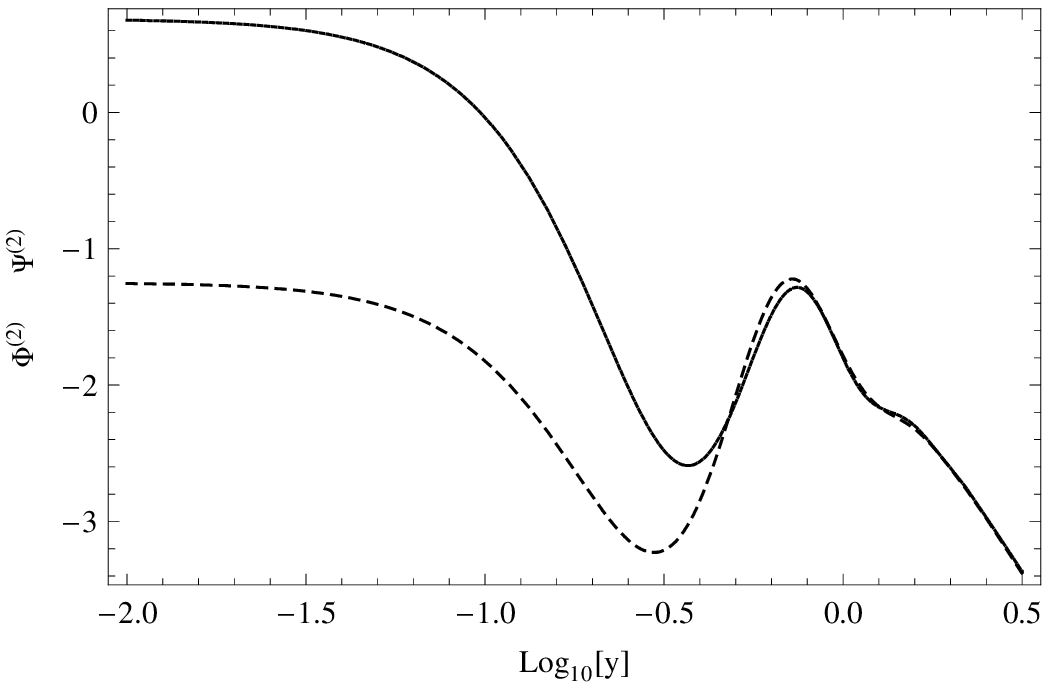}

\includegraphics[width=8cm]{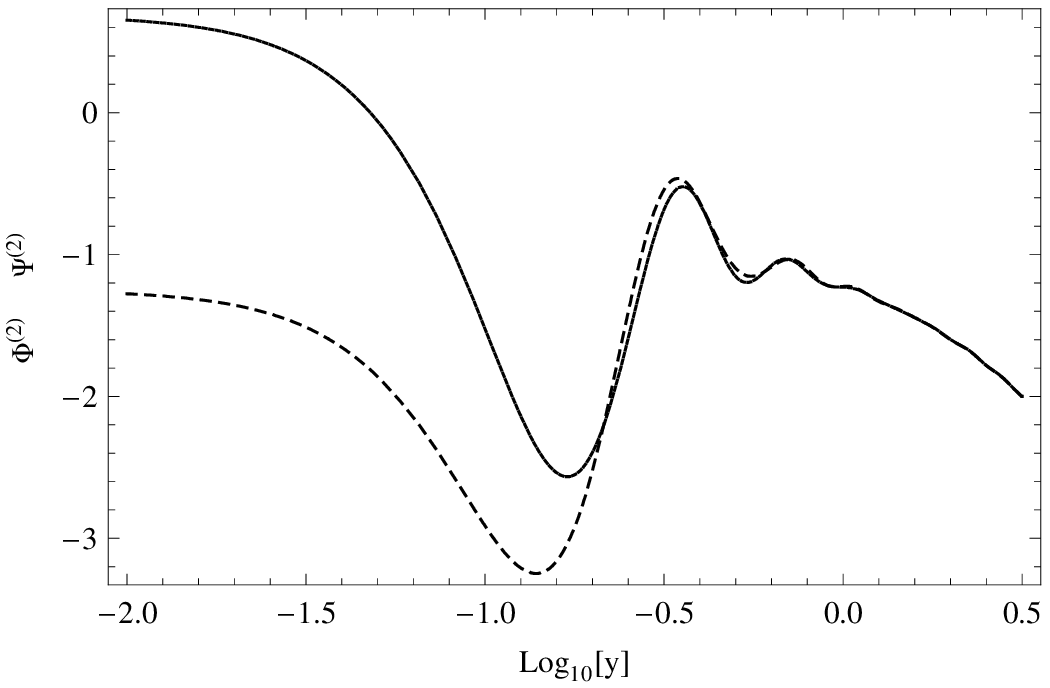}
\caption{Evolution of the two second order gravitational
potentials, $\phi^{(2)}(\vk_{1},\vk_{2})$ and $\psi^{(2)}(\vk_{1},\vk_{2})$ for $k_1=k_2= 10k_\EQ$
(top panel)
or $k_1=k_2=20k_\EQ$ (bottom panel) and $\vk_{1}\cdot\vk_{2}=0$.}
\label{figO2-1}
\end{figure}

\begin{figure}[htb]
\center
\includegraphics[width=8cm]{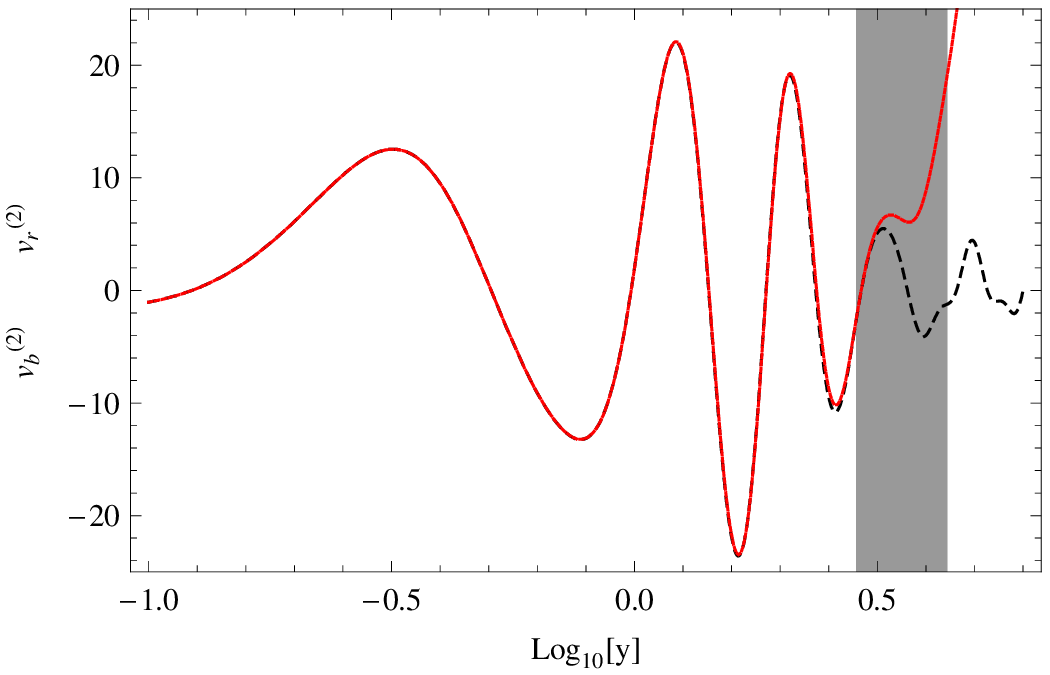}

\includegraphics[width=8cm]{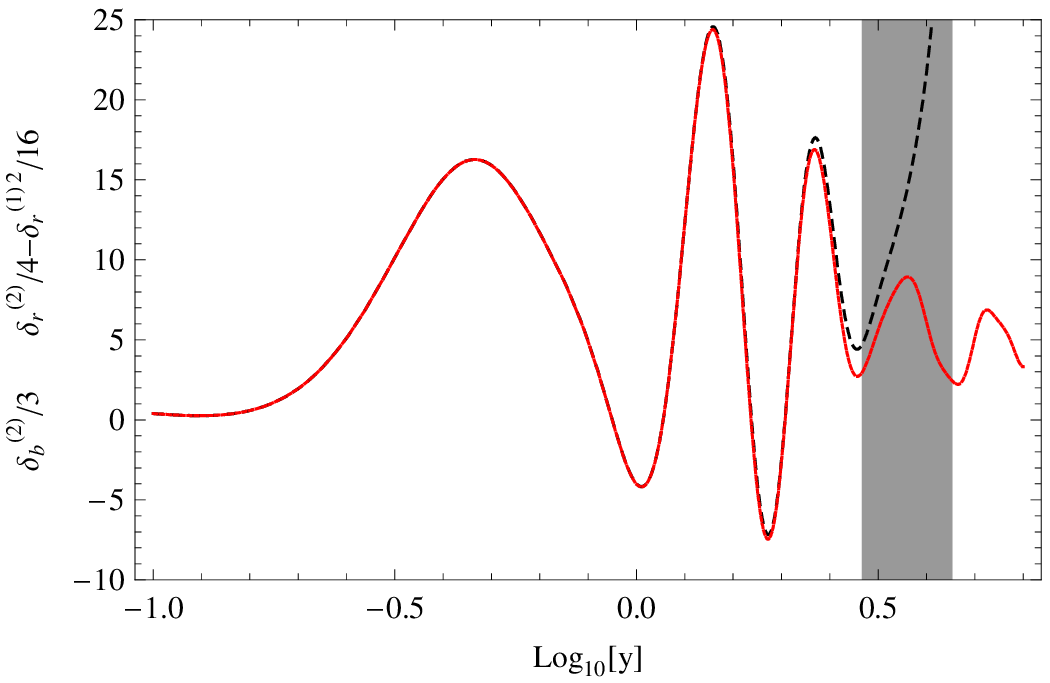}
\caption{Top: Comparison of the baryons and photons velocity
perturbation at order 2 for $k_1=k_2=10k_\EQ$ and $\vk_{1}\cdot\vk_{2}=0$. It
shows that $v_\rad^{(2)}=v_\baryon^{(2)}$ with a good
approximation until decoupling. Bottom: l.h.s and r.h.s of
Eq.~(\ref{CI2abiab}) for the adiabaticity condition at order 2. It
can be seen that this adiabaticity condition holds until
recombination, hence justifying the approximation of
\S~\ref{sec3a}.} \label{figO2-2}
\end{figure}

Fig.~\ref{figO2-1} shows the evolution of the two second order
gravitational potentials. In particular, it shows that the
solution is driven, as expected, toward $\Phi^{(2)}=\Psi^{(2)}$.
It is to be noted that the convergence takes place before
equivalence, at $y\equiv 1$, as stressed in the following. On the
other hand Fig.~\ref{figO2-2} depicts the evolution of the
velocities and density contrasts and shows that the photons-baryons
plasma can safely be described as a single fluid, almost until the
decoupling (shaded area on the figure).

\section{Analytical insight}\label{sec3a}

Before we proceed to describing the outcome of our numerical
integrations, and in order to gain some insight into the physics
of this intricate system, we present some analytic descriptions of
its solutions.

\subsection{Heuristic argument and hypothesis}\label{sec3a_heur}

Let us first assume that $f_\baryon\ll1$ so that the universe is
mainly dominated by non-interacting  cold dark matter and
radiation components. When, in the radiation era, the CDM
component is completely negligible the gravitational potential is
determined by the density contrast of radiation. The latter
however develops oscillations after Hubble-radius crossing while
those in the CDM fluid increases. It follows that, while still
formally in the radiation era ($\rho_{\rm r}>\rho_{\rm c}$), the
cold dark matter component is actually driving the gravitational
potential. It then acts as an external driving term in the evolution equation of radiation. In such a scenario we then
expect non-linearities that develop in the CDM sector to be
transferred first to the gravitational potential and then to the
radiation density fluctations.

To make this heuristic argument more quantitative we will thus assume that
\begin{itemize}
 \item we can first study the CDM-plasma system to determine
 the gravitational potential at second order where for simplicity the plasma
 is assumed to be radiation dominated;
 \item then study the acoustic oscillations of the baryon-photon
 plasma driven by the gravitational potential derived this way, both
 at first and second order in the perturbations.
\end{itemize}
For the sake of simplicity we will work in the tight coupling
approximation. We recall that the time of decoupling is both the
time at which this tight coupling regime ceases to be valid and
the time at which the radiation temperature is observed.

The tight coupling approximation amounts to saying that the
coupling terms $\mathcal{F}_\rad$ and $\mathcal{F}_\baryon$ are so
large that radiation and baryons behave as single fluid. It
ensures that the two fluids have the same peculiar velocity
($v_\rad=v_\baryon$) and implies that the anisotropic pressure of
radiation vanishes ($\pi_\rad=0$).

From Eq.~(\ref{Eqcons1}), it implies that
\begin{equation}\label{ratio}
 \delta_\baryon=\frac{3}{4} \delta_\rad\ .
\end{equation}
At linear order, eliminating $\mathcal{F}_a$ in
Eq.~(\ref{Eqcons2}) for radiation and baryons leads for the
photons-baryons plasma to the continuity equation
\begin{equation}\label{Eqcons2con}
 \delta_{{\rm pl}}'+ 3\HH (c^2_{s,{\rm pl}}-w_{{\rm pl}}) \delta_{{\rm pl}}+
 (1+w_{{\rm pl}})(\Delta v_{{\rm pl}}-3\Psi')=0\ ,
\end{equation}
and the Euler equation
\begin{equation}\label{Eqcons2pl}
v_{{\rm pl}}'+ \HH (1-3 c^2_{s,{\rm pl}}) v_{{\rm pl}}+ \Phi +
\frac{c^2_{s,{\rm pl}}}{1+w_{\rm pl}}\delta_{{\rm pl}} = 0\ ,
\end{equation}
where we have introduced the density contrast of the plasma
$\delta_{\rm pl}=\delta\rho_{\rm pl}/\rho_{\rm pl}$ with
\begin{equation}\label{deltapl}
 \delta\rho_{\rm pl}=\delta\rho_\rad+\delta\rho_\baryon\
 ,
 \quad
 \rho_{\rm pl}=\rho_\rad+\rho_\baryon\ .
\end{equation}
In the particular case at hand, it reduces to
$$
 \delta_{\rm pl}=\frac{1+R}{1+\frac{4}{3}R}\delta_\rad
$$
and the velocity perturbation are given by
$$
 v_{{\rm pl}}=v_\rad=v_\baryon\ .
$$
The equation of state and sound speed of the plasma are easily
obtained from the fact that $P_{{\rm pl}}=P_\rad$. They are
explicitly given by
\begin{equation}\label{wplasma}
 w_{\rm pl}=\frac{1}{3+4R}\ ,
 \quad
 c^2_{s,{\rm pl}}=\frac{1}{3(1+R)}\ ,
\end{equation}
and are time-dependent quantities (simply because the relative
contribution of the two components changes with time).

At second order, the density contrast and velocity perturbation of the plasma are given by
$$
 \delta^{(2)}_{\rm pl}=\frac{(1+R)\delta^{(2)}_\rad-\frac{R}{4}
   \delta_\rad^{(1)2}}{1+\frac{4}{3}R},\quad v^{(2)}_{{\rm pl}}=v^{(2)}_\rad=v^{(2)}_\baryon\ .
$$
It can be checked that the plasma follows the equation~(\ref{Eqcons2con})
and~(\ref{Eqcons2pl}) but supplemented with the source terms
\begin{eqnarray}
 S_{c,{\rm pl}} = S_{c,a= {\rm pl}}\, \qquad
 S_{e,{\rm pl}} = S_{e,a= {\rm pl}},
\end{eqnarray}
where $S_{c,a= {\rm pl}}$ and $S_{e,a= {\rm pl}}$ stand for
$S_{c,a}$ and $S_{e,a}$ in which we take the fluid to be the
plasma, that is $\delta_a=\delta_{\rm pl},\,\,w_a=v_{\rm pl}$, etc.

The validity of the tight coupling approximation at first and
second order can be checked from our numerical integration (which
indeed does not make this assumption) respectively on
Fig.~\ref{figO2-2} and Fig.~\ref{figO1-2} for the second and first
orders.

\subsection{CDM-radiation system}

\subsubsection{First order}\label{Sec_CDM_rad_1}

Deep in the radiation era, the gravitational potential is mainly
determined by the radiation density contrast and decays on
sub-Hubble scales. The contribution of matter is negligible in
the Poisson equation and it follows that the potential is given by
\begin{equation}\label{e.solPhiRDU}
 \Phi(k,\eta) = 3\Phi(k,\eta_\init)\frac{j_1(c_{s,\rad}x)}{c_{s,\rad}x},
\end{equation}
where $j_1$ is a spherical Bessel function of order 1 and
$x=k\eta$. The density contrast of radiation is given by
$\delta_\rad=-2\Phi$ on super-Hubble scales (see
\S~\ref{para.CI1}).

Let us now turn to the evolution of the CDM fluid during the
radiation era. In terms of the variable $y$ the continuity and
Euler equations lead to
\begin{eqnarray}
\ddot{\delta}_c+\frac{2+3y}{2y(1+y)}\dot{\delta}_c =  S_{\Phi}(y)\
,\label{e:dc}
\end{eqnarray}
with a driving force determined by the gravitational potential
\begin{eqnarray}
 S_{\Phi}(y) = 3\ddot{\Phi}+\left[\frac{6+9y}{2y(1+y)}
 \right]\dot{\Phi}-\frac{2}{1+y}\left(\frac{k}{k_\EQ}\right)^2\Phi\ ,
\label{e:dc2}
\end{eqnarray}
where a dot stands for a derivative with respect to $y$. The general
solution of Eq.~(\ref{e:dc}) is of the form
$\delta_c(k,\eta)=A+B\ln x +\delta_{\rm part}$ where $\delta_{\rm
part}$ is a particular solution given by
$$
 \delta_{\rm part} = \int_{\eta_\init}^\eta S_\Phi(k,\eta')\,
 \eta'\ln\left(\frac{\eta}{\eta'}\right)\dd\eta'\ ,
$$
where $S_{\Phi}$ is given by Eq.~(\ref{e.solPhiRDU}) as long as
$\delta\rho_\cdm\ll\delta\rho_\rad$. For $y\ll1$, the contribution
of the particular solution is negligible so that
$A\simeq-\frac32\Phi(\eta_\init)$ and $B\simeq0$. The
solution~(\ref{e.solPhiRDU}) shows that $\Phi$ vary mainly when
$\eta\sim k^{-1}$ so that $\delta_{\rm part}\sim A+B\ln x$. $A$
and $B$ can be obtained semi-analytically and are well
approximated by $A\simeq6$ and $B\simeq-9$ so that $\delta_\cdm
\simeq \Phi(k,\eta_\init)(-4.5 + 9 \ln x)$.

\begin{figure}[htb]
\center
\includegraphics[width=8cm]{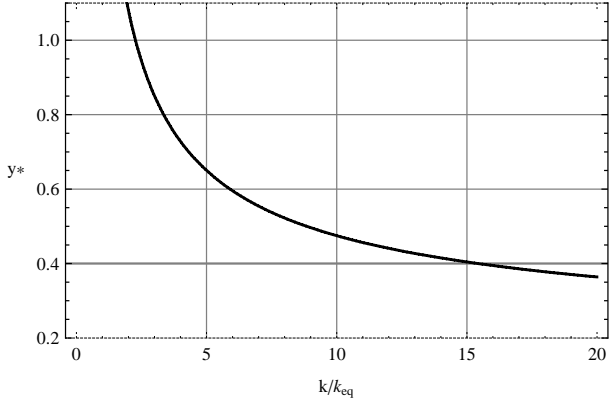}
\caption{The time at which the contribution of cold dark
matter and of radiation are comparable in the Poisson equation
as a function of $k/k_\EQ$. $k_\EQ$ is the wave-number of the mode
that becomes sub-Hubble at the time of quality. For most of the
scales of interest, i.e.  for $k\gg k_{\rm eq.}$, the contribution
of CDM in the Poisson equation is dominant before equality,
i.e. $y_\star<1$).} \label{ystar}
\end{figure}

This solution is valid as long as $\delta\rho_c\ll\delta\rho_r$ in
the Poisson equation. However, on sub-Hubble scales $\delta_\rad$
remains constant while, as we just saw, $\delta_c$ grows
logarithmically. Their contribution in the Poisson equation then
become to be of the same order when $y\sim y_\star(k)$, where
$y_\star$ is solution of $y_\star[-4.5 + 9 \ln(\sqrt{2}y_\star
k/k_\EQ)]\sim 6/(1-\fracb)$, where use has been made of
$x=\sqrt{2}k y/k_\EQ$ as long as $y\ll1$,. The solution of this
equation is depicted on Fig.~\ref{ystar}. For most of the scales
of interest, i.e.  for $k\gg k_{\rm eq.}$, the contribution of the
CDM in the Poisson equation is dominant before equality, i.e.
$y_\star<1$).

For these modes, which became sub-Hubble during the radiation era,
we shall consider that CDM dominates in the Poisson equation and
neglect the contribution of the radiation density perturbation, so
that the Poisson equation takes the form
\begin{equation}
 k^2 \Phi = -\frac{3(1-\fracb)}{4 y} k_\EQ^2 \delta_\cdm\ .
\end{equation}
Neglecting the contribution of baryons, since their density
contrast cannot grow because they are tightly coupled to the
radiation, Eq.~(\ref{e:dc}) then takes the form of the
M\'esz\'aros equation~\cite{Meszaros:1974}
\begin{equation}\label{Meszaros}
 \ddot\delta_\cdm + \frac{2+3y}{2y(1+y)} \dot\delta_\cdm -
 \frac{3}{2y(y+1)}\delta_\cdm = 0\ .
\end{equation}
Its two solutions are a growing mode
\begin{equation}
 D_+(y)=y+ 2/3
\end{equation}
and the decaying mode
\begin{equation}\label{solMeszaro}
 D_-(y) = -2\sqrt{(1+y)} + D_{+}(y)\ln\left(\frac{\sqrt{1+y}+1}{\sqrt{1+y}-1}
\right)\ .
\end{equation}

\subsubsection{Second order}

At second order, as at first order, while deep in the radiation
era the gravitational potential generated by the density contrast
of radiation decays when a mode becomes sub-Hubble. At this order
the gravitational potential satisfies
\begin{eqnarray}\label{Eq_Psi_ordre_2}
 \Psi''+4 \HH \Psi' -\frac{1}{3}\Delta \Psi&=& S_r\ ,
\end{eqnarray}
with
$$
 S_r = S_2
-\frac{1}{3}S_1+\frac{1}{3}\Delta S_3+\HH S_3'\ ,
$$
where the source terms are given by Eqs.~(\ref{e.S1}-\ref{e.S4}).
Up to a fast decaying solution, the general solution is
\begin{equation}
 \Psi(\eta)=3\Psi(k,\eta_\init)\frac{j_1(c_{s,\rad}x)}{c_{s,\rad}x}
  + \int_{\eta_\init}^{\eta}G_r(k,\eta,\eta')S_r(\eta')\dd \eta'\ ,
\end{equation}
with the Green function
\begin{eqnarray}
 G_r(k,\eta,\eta') &=& -\frac{\eta'}{c_{s,\rad}^3x^3}
     \Big\{\left(c_{s,\rad}^2xx'+1\right) \sin \left[c_{s,\rad}
    (x-x') \right]\nonumber\\
    && -c_{s,\rad}(x-x')\cos\left[c_{s,\rad}(x-x')\right]\Big\}\ .
\end{eqnarray}
On sub-Hubble scales, the leading terms in $S_r$ are those
quadratic in the first order velocity, of the form $\propto\HH^2
\dhi v \dbi v \sim \eta^{-2}$, which behave as $\eta^{-2}$. All
other terms in $S_r$ behave, at best, as $k^{-2}\eta^{-4}$. Using
the first order solution, the second order gravitational potential
asymptotically behaves as
\begin{eqnarray}
\Psi^{(2)}_S(k,\eta)&\simeq& -\frac{81}{2}\mathcal{C}
  \frac{\Psi^{(1)}(k_1,\eta_\init)\Psi^{(1)}(k_2,\eta_\init)}{k^2\eta^2(1-\mu^2)}\nonumber\\
&&\times\left[ 1 +2 \left(\frac{1}{k_1^2}+\frac{1}{k_2^2} \right)
  \gr{k}_1\cdot\gr{k}_2 +3\left( \frac{ \gr{k}_1\cdot\gr{k}_2}{k_1 k_2}
  \right)^2\right]\nonumber\\
&&\times\left[
  \cos\left(c_{s,\rad}k_1\eta\right)\cos\left(c_{s,\rad}k_2\eta\right)
  -\cos\left(c_{s,\rad}k\eta \right)\right.\nonumber\\
&&\qquad\qquad \left.-\mu \sin\left(c_{s,\rad}k_1\eta
\right)\sin\left(c_{s,\rad}k_2\eta\right) \right]\ ,
\end{eqnarray}
and it can be checked that this term is indeed regular in
$\mu^2=1$. Taking the homogeneous solution into account,
$\Psi^{(2)}$ decays as $(k\eta)^{-2}$ on sub-Hubble scales.

Now, the evolution of the density contrast of CDM follows, using
$y$ as the time variable, the evolution equation
\begin{eqnarray}\label{Eq_Evolution_deltam_2}
&&\ddot{\delta}_c^{(2)}+\frac{2+3y}{2y(1+y)}\dot{\delta}_c^{(2)}\\
&&-3\ddot{\Psi}^{(2)}-\left[\frac{6+9y}{2y(1+y)}
\right]\dot{\Psi}^{(2)}+\frac{2}{1+y}\left(\frac{
k}{k_\EQ}\right)^2\Phi^{(2)} = S_{\delta_c}\ ,\nonumber
\end{eqnarray}
where the source term is given by
$$
 S_{\delta_c} \equiv \frac{1}{k_\EQ}\sqrt{\frac{2}{1+y}}\left(\dot{S}_{c} +
  \frac{S_c}{y}\right)+\frac{2}{1+y}\left(\frac{k}{k_\EQ}\right)^2S_{e}\ .
$$
As in the previous section for the first order perturbations, CDM
density perturbations grow faster than those of radiation so
that, for any mode $k$ that became sub-Hubble before $\eta_\EQ$,
there exists a time of order $y_\star[k]$ such that for
$y>y_\star[k]$ the gravitational potential at second order is
determined by the cold dark matter. Whereafter, even though we are
still in the radiation era, we can neglect the contribution of the
density perturbation of radiation so that the Poisson equation
becomes
\begin{equation}\label{e.GGG}
 k^2 \Phi^{(2)} \simeq k^2 \Psi^{(2)}\simeq -\frac{3(1-\fracb)}{4
y} k_\EQ^2 \delta^{(2)}_c\ .
\end{equation}
In this regime, the evolution of the density contrast of CDM
at second order can be derived from a second order
M\'esz\'aros-like equation, in a similar way as at first order.
Using Eq.~(\ref{e.GGG}) and the fact that the main contributions
in $S_{\delta_c}$ in this regime come from
\begin{equation}
 S_{c} \simeq -2 \dbi\left( \delta \dhi v\right), \quad \dbi S_{e} \simeq -2\left(\dbj
  v \dhj \dbi v \right)\ ,
\end{equation}
Eq.~(\ref{Eq_Evolution_deltam_2}) takes the form
\begin{equation}\label{e.JKL}
\ddot{\delta}_c^{(2)}+\frac{2+3y}{2y(1+y)}\dot{\delta}_c^{(2)}
-\frac{3(1-\fracb)}{2y(1+y)}\delta_c^{(2)} = S_{M}\ ,
\end{equation}
with
\begin{eqnarray}
S_{M}&=&\mathcal{C}\Bigg\{\left[2 \delta_c \ddot \delta_c+2
  \dot \delta_c^2+\frac{\delta_c \dot \delta_c}{y(1+y)}
\right]+2\frac{\gr{k}_1\cdot\gr{k}_2}{k_1^2 k_2^2}\dot \delta_c^2
 \\
&&\hspace{-1cm}+\left[ \delta_c \ddot \delta_c+2
  \dot \delta_c^2+\frac{\delta_c \dot \delta_c}{2y(1+y)}
\right]\gr{k}_1\cdot\gr{k}_2\left(\frac{1}{k_1^2}+\frac{1}{k_2^2}
\right)\Bigg\}\, .\nonumber
\end{eqnarray}
Now we shall neglect the effect of baryons, that is
$f_\baryon$. This equation could then be called the second order
M\'esz\'aros equation and describes a growth of CDM density
perturbation in a regime where radiation dominates the dynamics of
the background while its density perturbations are negligible in
the Poisson equation.

The Green function associated to this equation is obtained to be
\begin{eqnarray}
 G(y,y') &=& \frac{3}{2}y'\sqrt{1+y'}(2+3y)(2+3y')\times \nonumber\\
         && \left[\frac{\sqrt{1+u}}{2+3u}
                   -\frac{1}{6}\ln\frac{\sqrt{1+u}+1}{\sqrt{1+u}-1}
                   \right]_{u=y'}^{u=y}\ ,
\end{eqnarray}
so that the general solution of Eq.~(\ref{e.JKL}) is
$$
 \delta_c^{(2)}=C_1D_+(y)+C_2 D_-(y)+\int_0^y G(y,y')S_M(y')\dd y'\ .
$$

\ifjesuisfrancis
\begin{figure}[htb]
\center
\includegraphics[width=8.5cm]{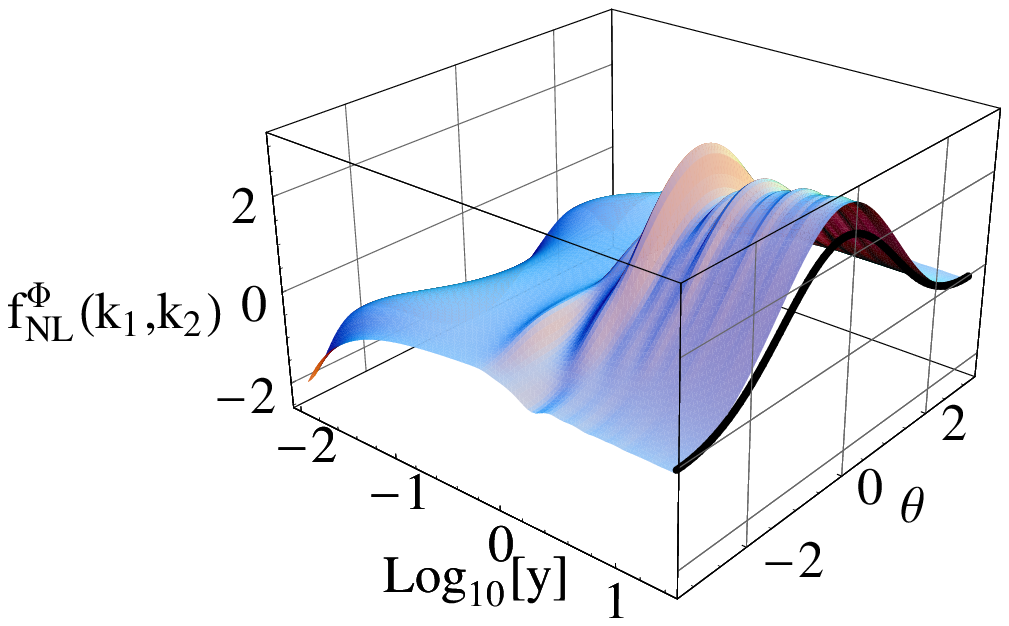}
\caption{(solid line) The second order potential computed in the
tight coupling limit as a function of time $y$ and of $\theta$,
angle between the wave vectors. It is compared to its expected
late time behaviour~(\ref{Phi2MD}). It is to be noted that the
convergence toward this solution is extremely rapid and takes
place as soon as equality is reached, e.g. $y=1$. The results
correspond to $k_{1}=6k_\EQ$ and $k_{2}=12k_\EQ$. The difference
in the amplitude of the function is due to the fact that the
baryons component has been neglected in the derivation of
Eq.~(\ref{Phi2MD})}. \label{Phi2Tight}
\end{figure}
\else\fi

In the limit where $y \gg 1,\, y'\gg 1$, the Green function
behaves as
$$
  G(y,y') \simeq \frac{2}{5}y \left[1-
 \left(\frac{y'}{y}\right)^{5/2}\right],
$$
and the source term as
$$
 S_M\simeq 7 \mathcal{C}  K(\gr{k}_1,\gr{k}_2)
 \frac{\delta(k_1)\delta(k_2)}{y^2}\ ,
$$
with the Kernel
\begin{equation}
K(\gr{k}_1,\gr{k}_2)\equiv\left[ \frac{5}{7} + \frac{1}{2}
\gr{k}_1\cdot\gr{k}_2
  \left(\frac{1}{k_1^2} + \frac{1}{k_2^2}\right)+ \frac{2}{7}
  \frac{(\gr{k}_1\cdot\gr{k}_2)^2}{k_1^2k_2^2}  \right]\ .
\end{equation}
In the limit $y\gg1$, the particular solution dominates and our
solution converges toward
\begin{equation}
 \frac{1}{2}\delta^{(2)}(k) \simeq \mathcal{C}K(\gr{k}_1,\gr{k}_2)
\delta_c(k_1)\delta_c(k_2)\ ,
\end{equation}
that is toward the standard result~(\ref{F2Peeb}) describing the
collapse of cold dark matter in a matter dominated era. The second
order gravitational potential is then obtained from the Poisson
equation
\begin{equation}\label{Phi2MD}
\frac{1}{2}\Phi^{(2)}(k,\eta) \simeq  - \mathcal{C}
\frac{1}{6}K(\gr{k}_1,\gr{k}_2)\left(\frac{k_1
    k_2 \eta}{k}\right)^2 \Phi(k_1)\Phi(k_2),
\end{equation}
up to terms of order $\mathcal{O}(f_\baryon)$. The convergence
towards the solution~(\ref{Phi2MD}) is explicitly depicted on
Fig.~\ref{Phi2Tight} where the behaviour of the exact (numerically
integrated) second order potential as a function of time (and
angle) is compared to its expected late time
behaviour~(\ref{Phi2MD}). As detailed above, this solution is a
better approximation for larger wave-numbers and at large $y$
since we converge to this solution for $y>y_\star(k)$. On this
figure it can be observed that the convergence is extremely rapid
and that the full kernel structure, including its angular
dependence in (\ref{Phi2MD}), is indeed to be observed in
$\Phi^{(2)}$.

\subsection{Baryons-radiation system}

We now want to understand the behaviour of the baryons-photons
plasma, and in particular of its acoustic oscillation, in the
regime in which the gravitational potential is determined by the
solutions of the previous section.

We restrict our analysis to the tight coupling regime. And since
it occurs for $y<y_\LSS$, our solution will gain in accuracy when
the period between CDM domination, $y= y_\star(k)$, and the last
scattering surface, $y=y_\LSS$, is large, that is on the smallest
scales.

\subsubsection{First order}

The computation at first order is well
known~\cite{1996ApJ...471..542H} and we review its main steps to
compare to the more unexplored second order case.

In the fluid limit, the baryons and photons both obey a continuity
and conservation equations~(\ref{Eqcons1}-\ref{Eqcons2}) with a
source term and in which the anisotropic stress of radiation
can be neglected because of the tight-coupling approximation. As
discussed in \S~\ref{sec3a_heur}, in the tight coupled regime,
that is for $k/\tau'\ll 1$, it leads to the wave equation
\begin{eqnarray}\label{tight}
 &&\left(\frac{\delta_\rad}{4}- \Psi\right)''
 +\frac{R'}{1+R}\left(\frac{\delta_\rad}{4}- \Psi\right)' + k^2c_s^2
 \left(\frac{\delta_\rad}{4}- \Psi\right) \nonumber\\
&& \qquad\qquad \simeq \frac{-k^2}{3}\left[\Phi\left(1+
\frac{1}{1+R}
  \right)\right]\ ,
\end{eqnarray}
where $R$ is defined in Eq.~(\ref{e.defR}), the sound speed is
given by Eq.~(\ref{wplasma}). This is a wave equation with a
forcing term on the r.h.s. which describes the oscillations of the
plasma.

For small wavelength modes, the variation of $R$ and $\Phi$ is
small compared to the period of the wave so that we can construct
an adiabatic solution by resorting on a WKB approximation; see
e.g. Ref.~\cite{1996ApJ...471..542H} for details. Defining
$$
\Theta_{\text{SW}} \equiv \frac{1}{4}\delta_\rad + \Phi
$$
and the sound horizon
$$
 r_s(\eta) \equiv \int_0^{\eta}\frac{1}{\sqrt{3[1+R(\eta')]}} \dd
 \eta'\ ,
$$
the WKB solution takes the form
\begin{equation}
\Theta_{SW}(k, \eta)=\frac{\left[ \Theta_{SW}(0)+R\Phi
\right]}{(1+R)^{1/4}}\cos[k r_s(\eta)] -R \Phi
 \ .
\end{equation}
The velocity field, $v_\rad=v_\baryon$ can then be determined from
the Euler equation~(\ref{Eqcons2pl}).

This solution neglects the Silk damping effect that can be
described by adding terms in $k^2/\tau'$ in Eq.~(\ref{tight}) so
that the solution is exponentially suppressed by a factor
$$
 \mathcal{D}(k,\eta) =\exp\left(-\frac{k^2}{k_D^2}\right)
$$
where
$$
 k_D^{-2}(\eta)\sim\frac{1}{6}\int_0^{\eta}
\frac{1}{1+R(\eta')}\left[\frac{16}{15}+\frac{R^2(\eta')}{1+R(\eta')}\right]\frac{\dd
\eta'}{\tau'(\eta')}\ ,
$$
so that
\begin{equation}\label{e.73}
\Theta_{SW}+R \Phi\propto \exp\left({-\frac{k^2}{k_D^2}}\right)
 \ ,
\end{equation}
where the damping scale is of the order of $k_D\sim 15 k_\EQ$.
Since $\Phi$ decreases as $(k/k_\EQ)^{-2}$, we conclude that for
large wavenumber, $\Theta_{SW}(k, \eta)$ goes rapidly to zero.

\subsubsection{Second order}\label{sec.O2.analytic}

The former approach can be generalised at second order, but the
behaviour of $\Theta_{SW}$ will change mainly because the second
order version of Eq~(\ref{tight}) has a r.h.s. which is steadily
growing in the range of interest, i.e. after $\eta_\EQ$ and on
large $k$.

At second order, Eq.~(\ref{tight}) will also contain terms coming
from the second order Liouville equation~\cite{Pitrou:2007jy} of the form
$$
S_{\rm pl}= \frac{1}{4}\left( S_{c,\rad}'+
  \frac{R'}{1+R}S_{c,\rad}\right)+\frac{k^2}{3}S_{e,{\rm pl}}\,.
$$
This source term involves terms which are quadratic in the fluid
perturbation variables ($\delta,v$) and the potentials
($\Phi,\Psi$). The former are exponentially suppressed due to Silk
damping and the latter decrease as $(k/k_\EQ)^{-2}$. We can thus
neglect this source term as long as we focus on small scales.
Defining,
\begin{equation}\label{defSW2}
\Theta_{SW}^{(2)}\equiv \frac{\delta_\rad^{(2)}}{4} + \Phi^{(2 )}\
,
\end{equation}
the solution for $\Theta_{SW}^{(2)}$ will be similar to
Eq.~(\ref{e.73}). When Silk damping is taken into account,
$\Theta_{SW}^{(2)}+R\Phi^{(2)}$ is exponentially suppressed,
exactly as at first order. The main difference with first order
arises from the fact that the second order gravitational
potentials are driven toward $\Psi^{(2)} \simeq \Phi^{(2)} \sim
(k\eta)^2$ after equivalence ($y>y_\EQ$) so that we expect that
\begin{equation}\label{SW2}
\Theta_{SW}^{(2)}\equiv \frac{\delta_\rad^{(2)}}{4} + \Phi^{(2 )}
\simeq - R\Psi^{(2)}\ .
\end{equation}
Now, the velocity of radiation is given by
\begin{equation}
 \left[(1+R)v_\rad^{(2)}\right]' = -\frac{\delta^{(2)}_\rad}{4} - (1+R) \Phi^{(2)}\ .
\end{equation}
Since $\Theta_{SW}^{(2)}+R\Phi^{(2)}$ is exponentially suppressed
due to the Silk damping, we expect that the Doppler contribution
is also negligible.

The equations (\ref{Phi2MD}) and (\ref{SW2}) are the central
results of the analytic insight of the non-linear regimes of the
photons-baryons-CDM system at second order. They give the behaviour
of the second order gravitational potentials at the time of
decoupling together with the response of the photon-electron
plasma. We also conclude that we expect
$\Theta_{SW}^{(2)}=-R\Phi^{(2)}$ to dominate the CMB temperature
anisotropies on small angular scales (see Appendix~\ref{appB} for
a discussion of the integrated Sachs-Wolfe contribution).

The bottom line of our analytic estimates is that on small angular
scales, i.e. $k/k_\EQ\gg1$, the density perturbation of CDM starts
to dominate the Poisson equation from $y_\star(k)$ so that from
this time to decoupling we can assume that the system is split
in (1) the evolution of CDM and (2) the evolution of the
photons-baryons plasma which develops acoustic oscillations in the
gravitational potential determined by the CDM component. Because of Silk
damping, $\Theta_{SW}+R\Phi$ dies out on small scales. At first
order this implies that $\Theta_{SW}^{(1)}\sim-R\Phi^{(1)}$ which
is suppressed by a factor $k_\EQ^2/k^2$ due to its evolution in
the radiation era prior to $y_\star(k)$. At second order however,
we still have for the same reason that
$\Theta_{SW}^{(2)}\sim-R\Phi^{(2)}$ but now this term roughly
grows as $(k\eta)^2$. Note that since $k_D$ is of order $15k_\EQ$
and that $k_\EQ$ roughly corresponds to a multipole $\ell\sim160$,
we expect our analysis to give a good description of the system
for $\ell\gtrsim2400$.

\subsection{Comparison to numerics}\label{sec-num-an}

We now turn to the description of the numerical solutions of the
system described in \S~\ref{Integrations}. Its solutions will be
described in the light of the analytic description we just
developed.

\begin{figure}[htb]
\center
\includegraphics[width=8cm]{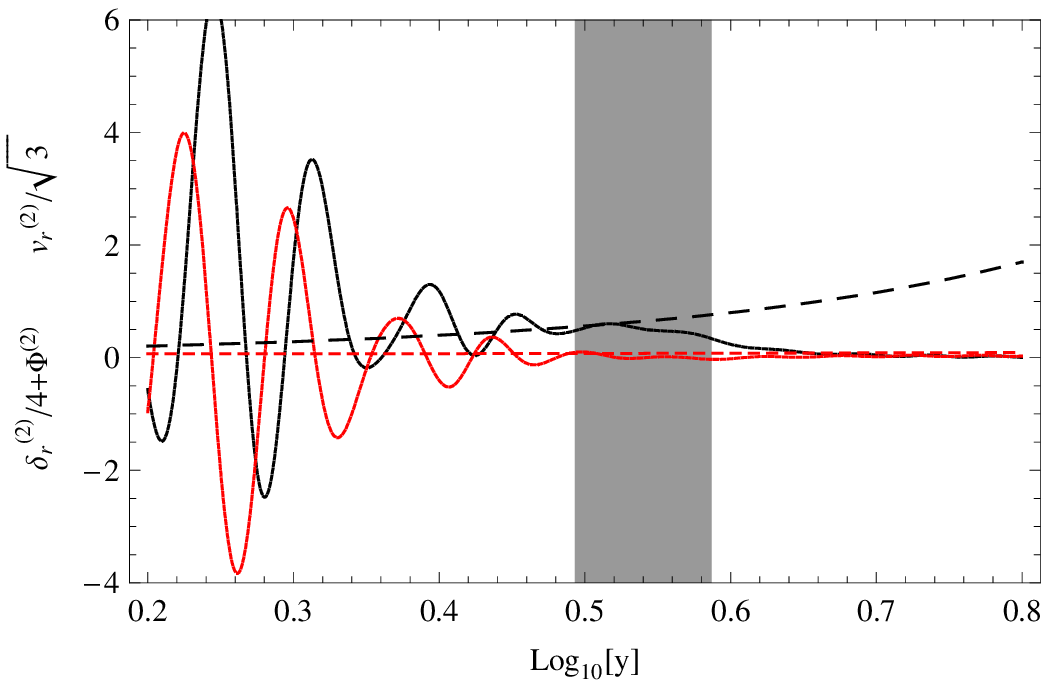}
\includegraphics[width=8cm]{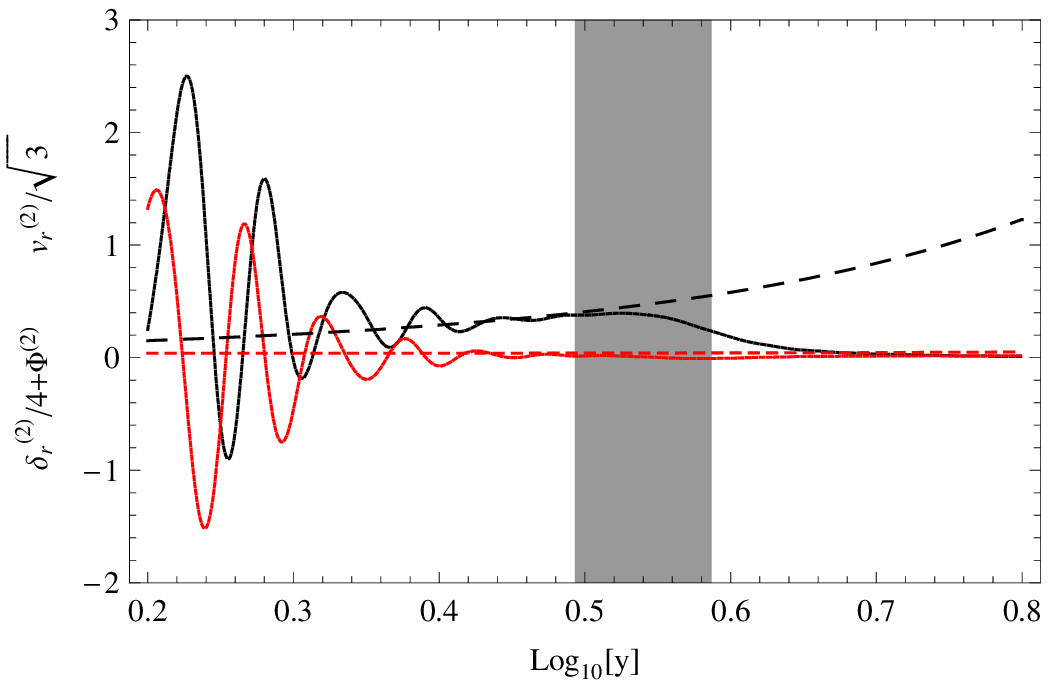}
\caption{Behavior of $\Theta_{SW}^{(2)}$ (black) and $k
v_\rad^{(2)}/\sqrt{3}$ (red). The numerical integration is
depicted with solid lines while the analytical estimates is
plotted in dashed lines. From top to bottom, we have
$k_{1}=k_{2}=30\,k_{\rm eq.}$ and $k_{1}=k_{2}=40\,k_{\rm eq.}$.
The vertical grey zone represents the ``surface'' of last
scattering.} \label{SW2Fig}
\end{figure}

Fig.~\ref{SW2Fig} shows the result of numerical integrations for
second order quantities of interest. They are compared  to our
approximate formula that, we recall, is expected to be valid in
the tight coupling regime. It shows indeed that modes with
$k>k_\EQ$ relax temporarily toward the solution~(\ref{SW2}). The
exact solution exhibits  though large oscillations that are
thought to be due to the acoustic oscillations that are present in
the plasma at first order, but their average turns out to coincide
with the proposed analytic formula (long dashed lines) as long as
a strong coupling is ensured. The Silk damping effect is observed
to play a key role to actually damping the oscillations. This
effect is all the more important that $k$ is large. Note that the
impact of the oscillations on the observational quantities will
also be damped by the finite width of the last scattering surface.
The wave-number corresponding to this width is of order of
$10k_\EQ$ which is smaller than the damping scale.

When the coupling becomes loose, $\log_{10}(y)$ approaching $0.6$,
the numerical solution departs from the expected solution (and it
converges toward 0) as the full Boltzmann hierarchy is now at
play. We also depict the velocity term which can be checked to be
negligible, as expected.

From these set of results we can then argue that the approximate
analytic solution described in (\ref{SW2}) captures the physics of
the dominant terms of the CMB anisotropies on small angular
scales. Here we have explicitly checked that this form is
consistent with the physics of recombination when the collision
effects are taken into account. We limit though the collision
effects to their first order expression. We expect nonetheless
that an exact calculation, up to second order, would not
significantly alter our conclusions, the collision physics playing
a role only during a limited period of time. Although this is
certainly desirable to do such a calculation, this is beyond the
scope of this paper whose goal is to estimate the order of
magnitude of the non-Gaussianity on these scales.

In the following we explore the observational consequence of such
a finding on the temperature bispectrum at small scale.

\section{Signature in the cosmic microwave background}\label{sec3}

\subsection{Flat sky approximation}\label{sec4.1}

Since our approximations hold on small angular scales, it is amply
sufficient to treat the sky as flat to compute the properties of
the CMB anisotropies. We thus decompose the CMB temperature
anisotropies in 2D-Fourier space as
\begin{equation}
 \Theta(\gr{n}) = \int \frac{\dd^2 \gr{l}}{2 \pi}
  \Theta(\gr{l})\, \hbox{e}^{\textrm{i} \gr{l}\cdot\gr{n}}\ ,
\end{equation}
so that
\begin{equation}
 \Theta(\gr{l}) = \int \frac{\dd^2 \gr{n}}{2 \pi}\Theta(\gr{n})
  \, \hbox{e}^{\textrm{-i} \gr{l}\cdot\gr{n}}\ .
\end{equation}
On the other hand $\Theta(\gr{n})$ can be expanded in Fourier
modes as
\begin{equation}\label{thetaden}
 \Theta(\gr{n})=\int \frac{\dd \gr{k}}{(2\pi)^{3/2}}
  \bar\Theta(\gr{k},\eta_\LSS)\hbox{e}^{\textrm{i}(k_r \eta_\LSS +
  D_\LSS\gr{k_{\perp}}\cdot\gr{n})}\ ,
\end{equation}
where $D_\LSS$ is the angular distance of the last scattering
surface given by $D_\LSS=\int_{y_{LSS}}^{y_0} 1/(H[u]u)\dd u$.
$\gr{k}_\perp$ is the projection of $\gr{k}$ on the sky, \i.e.
$\gr{k}=k_r \gr{e}+\gr{k_\perp}$ where $\gr{e}$ is the direction
of the (flat) sky.

In Eq.~(\ref{thetaden}) we refer to $\bar\Theta(\vk,\eta_\LSS)$ as
\begin{equation}\label{thetabar}
 \bar\Theta(\vk,\eta_\LSS)=\int \dd \eta
 \Theta(\gr{k},\eta)v(\eta,\eta_\LSS)\ ,
\end{equation}
considering the observed CMB anisotropies as a superposition of
spheres of temperature anisotropy weighted by the visibility
function $v(\eta,\eta_\LSS)$ which peaks at $\eta_\LSS$. The
angular power spectrum is nothing but the two dimensional power
spectrum of $\Theta(\gr{l})$,
\begin{equation}
 \langle\Theta(\gr{l})\Theta(\gr{l}')\rangle=
  \delta^{(2)}(\gr{l}+\gr{l}')C_{l}\ .
\end{equation}

We now need to determine $\Theta(\vk,\eta)$ in terms of the
perturbation variables. The CMB temperature anisotropies are
usually split as an intrinsic Sachs-Wolfe effect, a Doppler
effect and an integrated Sachs-Wolfe contribution. As discussed in
Appendix~\ref{appB} the integrated Sachs-Wolfe contribution is
expected to be negligible on the scales of interest in our study.

At first order, the Fourier component of temperature anisotropy
for a mode $\gr{k}$, in a direction $\gr{e}$ emitted at a comoving
distance $\eta_0-\eta$ is dominated by the Sachs-Wolfe and Doppler
terms,
\begin{eqnarray}
\Theta^{(1)}(\gr{k},\eta) &=& \Theta^{(1)}_{SW}(\gr{k},\eta)
           +\Theta^{(1)}_{\rm Dop}(\gr{k},\eta)\nonumber\\
&\equiv& g^{(1)}(\gr{k},\eta)\Phi^{(1)}_k(0)\ .
\end{eqnarray}
The Sachs-Wolfe term is related to the perturbation variables by
\begin{equation}
\Theta^{(1)}_{SW}(\gr{k},\eta)=\frac{\delta^{(1)}_r}{4}(\gr{k},\eta)+\Phi^{(1)}(\gr{k},\eta)
\ ,
\end{equation}
while the Doppler term is given $\Theta_{\text{Dop}}\equiv - v
\gr{k}.\gr{e}$ so that
\begin{eqnarray}
\Theta^{(1)}_{\rm Dop}(\gr{k},\eta)&=& \textrm{i}k_r
v^{(1)}_r(\gr{k},\eta)\ ,
\end{eqnarray}
with $k_r=\vk\cdot\gr{e}$. Using Eq.(\ref{thetaden}), we deduce
thus that
\begin{eqnarray}
 \Theta^{(1)}(\gr{l}) &=&\frac{1}{\sqrt{2 \pi}D_\LSS^2}\int \dd k_r  \hat
 g^{(1)}(\gr{k})\hbox{e}^{\textrm{i} k_r \eta_\LSS}\ ,
\end{eqnarray}
with
\begin{equation}
 \hat g^{(1)}(\gr{k})= \int \dd \eta v(\eta) g^{(1)}(\gr{k},\eta)
 \Phi^{(1)}(\vk,0)\
\end{equation}
and where $\vk_\perp= \gr{l}/D_\LSS$. Using the definition of the
initial power spectrum $\langle\Phi(\vk,0)\Phi(\vk',0)\rangle =
\delta^{(3)}(\gr{k}+\vk')P(k)$, we finally get that $C_l$ is given
by
\begin{equation}\label{Cl}
 C_l \simeq \frac{1}{2 \pi D_\LSS^2} \int \dd k_r
 P\left(\sqrt{k_r^2+\frac{l^2}{D_\LSS^2}}\right) \left|\hat
 g^{(1)}(\gr{k})\right|^2.
\end{equation}
This reproduces the main features of the CMB angular power
spectrum, as checked on Fig.~\ref{Cleps}.

At second order, and for the scales of interest ($k \gg k_{eq}$
and $k> k_D$), we stress again that the knowledge of the exact
expression of the terms quadratic in the first order variables in
the second order Sachs-Wolfe effect are not needed since they are
suppressed because of the Silk damping or because of the decaying
of the potential during the radiation era. The analysis of
\S~\ref{sec.O2.analytic} shows that the Doppler term is much smaller
than the intrinsic Sachs Wolfe term of Eq.~(\ref{SW2}). Thus, the
main contribution to the second order temperature anisotropy is
well approximated by
\begin{eqnarray}
\Theta^{(2)}(\vk,\eta)&\simeq& \Theta^{(2)}_{SW}(\vk,\eta) \ , \nonumber\\
 &\simeq& \frac{\delta^{(2)}_\rad(\vk,\eta)}{4}+\Phi^{(2)}(\vk,\eta) \simeq -R \Phi^{(2)}
 \ .
\end{eqnarray}
We deduce that
\begin{eqnarray}
 \Theta^{(2)}(\gr{l}) &=&\frac{1}{\sqrt{2 \pi}D_\LSS^2}\int \dd k_r  \hat
 g^{(2)}(\gr{k})\hbox{e}^{\textrm{i} k_r \eta_\LSS}\ ,
\end{eqnarray}
with
\begin{equation}
 \hat g^{(2)}(\gr{k})=\int \dd
 \eta\,v(\eta)\,\Theta^{(2)}(\vk,\eta).
\end{equation}
It can be rewritten in terms of the initial first order
gravitational potential as
\begin{equation}
 \hat g^{(2)}(\gr{k})=2\mathcal{C}\ \fNL^{(\Theta)}(\vk_{1},\vk_{2})
\Phi^{(1)}(\vk_1,0) \Phi^{(1)}(\vk_2,0)\ ,
\end{equation}
hence defining $\fNL^{(\Theta)}$.

\subsection{Bispectrum}\label{sec4.2}

In the flat sky approximation, the reduced bispectrum $b_{l_1 l_2
l_3}$ is defined from the 3-point function as
\begin{equation}
\langle\Theta({\bf l}_1)\Theta({\bf l}_2)\Theta({\bf
l}_3)\rangle=(2 \pi)^{-1}\delta^{(2)}({\bf l}_1+{\bf l}_2+{\bf
l}_3)b_{l_1l_2l_3} \ ;
\end{equation}
see e.g. Ref.~\cite{2002astro.ph..6039K}. With the previous
definitions, it can be expressed as
\begin{widetext}
\begin{eqnarray}
b_{l_1l_2l_3} &=& \frac{1}{2 \pi D_\LSS^4}
 \int \dd k_{r1} \dd k_{r2}
 \left[\fNL^{(\Theta)}(-\gr{k}_{1},-\gr{k}_{2})\hat g^{(1)}(\gr{k}_{1}) \hat g^{(1)}(\gr{k}_{2})
 P\left(\sqrt{k_{r1}^2+\frac{\vk_{\perp1}^2}{D_\LSS^2}}\right)
 P\left(\sqrt{k_{r2}^2+\frac{\vk_{\perp2}^2}{D_\LSS^2}}\right)\right]\nonumber\\
&&\qquad\qquad +\left(\gr{l_1}\rightarrow
\gr{l_2}\rightarrow\gr{l_3}\rightarrow
\gr{l_1}\right)+\left(\gr{l_1}\rightarrow
\gr{l_3}\rightarrow\gr{l_2}\rightarrow\gr{l_1}\right)\ .
\label{blllExp}
\end{eqnarray}
\end{widetext}

\subsection{Numerical computation}\label{sec4.3}

In order to perform the previous integrals, we need to specify the
initial power spectrum. We assume that the power spectrum is scale invariant
and we normalize it using the results of WMAP, that is
\begin{equation}
P(k) = 2\pi^2\frac{25}{9}A_s^2
\left(\frac{k_\EQ}{k}\right)^3\frac{1}{k_\EQ^3}
\end{equation}
with
\begin{equation}
A_S^2=3.33\times 10^{-10}\ .
\end{equation}
We then compute the bispectrum of an equilateral configuration for
which all momentums are equals; $l_1=l_2=l_3$. The only free
parameter for such configuration is the norm $l$ of the three
vectors.

\begin{figure}[htb]
 \includegraphics[width=10.5cm]{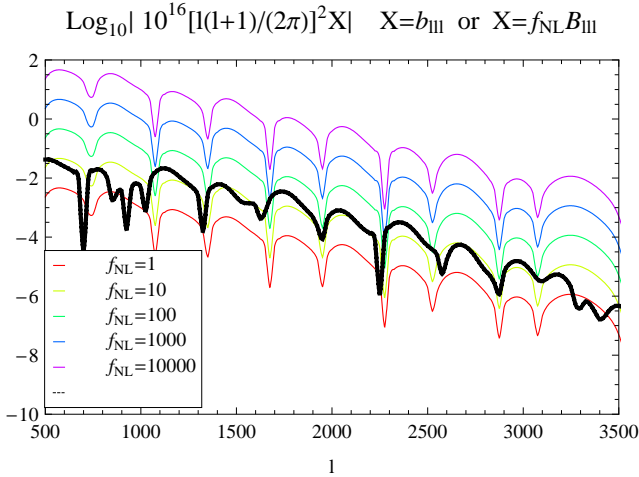}
  \caption{Black: the bispectrum for the equilateral configuration
  computed in the flat sky limit. The thin colored lines represents
  the bispectrum that would be obtained by assuming a
  constant initial $\fNL^{\Phi}$ and a linear transfer function,
  that is neglecting the non-linear dynamics. From bottom to top
  we have plotted $\fNL^{\Phi}=1,10,100,1000,10000$.}\label{figbispec}
\end{figure}

The result is depicted on Fig.~\ref{figbispec} and is compared to
the bispectrum one would obtain from a initial constant
$\fNL^{\Phi}$ assuming a linear transfer function. It appears that
on scales that range from $l=1000$ to $l=3000$ the bispectrum
resembles that of an effective constant primordial $\fNL^{\Phi}$
of order 25.

The order of magnitude of the amplitude of the bispectrum can be
understood from the following rule of thumb for modes larger than
$k_D$. According to our analysis, considering the equilateral
configuration where $k_1=k_2=k$, the second order temperature
anisotropy on the last scattering surface is of order
\begin{eqnarray}
\frac{1}{2}\Theta^{(2)}(\vk,\eta_\LSS)
   &\simeq& -\frac{1}{2}R_\LSS \Phi^{(2)}(\vk,\eta_\LSS)\nonumber\\
   &\simeq&-R_\LSS\frac{2y_{\LSS}}{3}
     \frac{k^2}{k_\EQ^2}\left[\Phi^{(1)}(k,0)\mTt^{(1)}(k)\right]^2
     \nonumber
\end{eqnarray}
where we have assumed that, on average, the Kernel
$K(\gr{k}_1,\gr{k}_2)$ is of order unity. Now, assuming a constant
primordial $\fNL$ evolved with the linear transfer function, the
second order temperature anisotropy would roughly be of order
\begin{equation}
\frac{1}{2}\Theta^{(2)}(\vk,\eta_\LSS)\simeq-R_\LSS f_{\text{NL}}
\left[\Phi^{(1)}(k,0)\right]^2 \mTt^{(1)}(k)
\end{equation}
since, for these modes, the integral on the visibility function
keeps only the average of the Sachs-Wolfe contribution. The ratio
of the contribution of the non-linear dynamics compared to a
primordial non-Gaussianity is
\begin{equation}
\frac{2}{3} \frac{y_\LSS}{\fNL} \left(\frac{k}{k_\EQ}\right)^2
\mTt^{(1)}(k)\ .
\end{equation}
Since for large modes the gravitational potential has been
decaying as $(k \eta)^{-2}$ in the radiation dominated era, and
growing logarithmically when the potential started to be
determined by the cold dark matter component(see the analysis of
section~\ref{Sec_CDM_rad_1}), we deduce that the first order
transfer function is typically given by
\begin{equation}
 \mTt^{(1)}(k)\simeq A(k)\left(\frac{k_\EQ}{k}\right)^2\ ,
\end{equation}
where $A(k)$ is a steadily growing function. At the Silk damping
scale we find numerically $A(k_D)\simeq 10$. We thus conclude that
in the bispectrum, the evolution for $l \simeq k_D D_\LSS \simeq
2400 $ is equivalent to a primordial
$$
 \fNL\simeq \frac{2}{3} y_\LSS A(k_D)\simeq 25
$$
evolved linearly. This estimates of the order of magnitude is in
complete agreement with Fig.~\ref{figbispec} where it can be read
that for multipole ranging from 2000 to 3000 the amplitude of the
bispectrum is comparable with the one that would be obtained from
a constant $\fNL$ ranging between 10 and 50.

There is no guarantee however that for arbitrary geometries the
shape dependence of the temperature bispectrum would be that of a
constant $\fNL$. It is rather determined by the kernel shape of
the form (\ref{SW2}).

\section{Conclusion}\label{sec5}

This article investigates the non-Gaussianity that arises in the
CMB temperature anisotropies due to the post-inflationary
non-linear dynamics during the radiation and matter dominated era.
More specifically it aims at identifying the leading mechanisms
and leading terms that determine the shape of the CMB bispectrum
on small angular scales.\\

The driving idea that we have pursued throughout the paper is that
at small angular scales the second order CMB anisotropies trace
the second order gravitational potential as it is shaped by the
CDM component during its sub-Hubble evolution. To give support to
this picture, we have developed both analytical insights into the
joint evolution of the density potentials and the temperature
fluctuations and numerical tools.

We have thus solved numerically the joint evolution equations of
the cosmic fluids up to second order. We have been able to check
that at the time of decoupling, the second order potential indeed
traces its expected shape. This conclusion is summarised and
illustrated on Fig.~\ref{Phi2Tight}. At this stage, and as long as
one restricts these results to the tight coupling regime, no
approximations have been made. The accuracy with which the
$k$-dependence of the matter dominated mode coupling kernel, i.e.
Eq.~(\ref{F2Peeb}), is recovered is truly remarkable. We stress
that this is due to the fact that for the physics at work at small
scales, i.e. $k/k_\EQ\gg1$, the density perturbations of the CDM
component start to dominate the Poisson equation much before
equality. This implies that the non-linearities developed by the
CDM can be transferred very efficiently to the gravitational
potential even before the beginning of the matter era.

Determining exactly how this mode coupling kernel is actually
transferred to the source term of the CMB anisotropies relies on
further numerical integrations through the recombination era. At
this stage, the {\it only approximation} we make concerns the
Compton scattering collision term entering the Boltzmann equation
for radiation at second order. We did not use its full second
order expression but we argue that it can be reduced to its formal
first order form, namely to Eq.~(\ref{e.Hyp}). Such an assumption
is clearly valid in the tight coupling regime. Actually this is
the only term appearing in the collision term in the baryons
rest-frame as long as tight coupling at first order is efficient.
We then argue that when the coupling drops, Silk damping effects
effectively suppress all other contributions.

This leads to the behaviour depicted on Fig.~\ref{SW2Fig} for the
main source term of the temperature anisotropies. In particular
$\Theta_{SW}^{(2)}$, the monopole of the second order source term,
is found to be attracted toward a non-vanishing and
non-oscillatory term, $-R\Phi^{(2)}$, where we recall that $R$ is
the baryon to radiation ratio, while the dipole contribution, and
thus the Doppler effect, vanishes. As a result the main
contribution to the CMB temperature anisotropies at second order
is found to be directly proportional to the second order
gravitational potential. It has to be noted that the efficiency
with which the second order term converges to this form is
considerably accelerated by the Silk damping effects which
efficiently suppress the oscillatory parts of the solution. This
results are clearly illustrated on Fig.~\ref{SW2Fig}. We argued
from order of magnitude arguments that, because the damping scale
is typically of order $15k_\EQ$, our description
shall be valid for $l\gtrsim2400$. This observation is the basis
of the main result of this paper. Actually,as Fig.~\ref{figbispec}
tends to show, it seems that this description may be valid at
lower multipoles.

Finally we explore the consequence on the CMB bispectrum. For
obvious reasons we use the small angle approximation to perform
the numerical integrations. The bispectrum for equilateral
configurations is illustrated on Fig.~\ref{figbispec}. We show
that for these configurations, its amplitude corresponds to what a primordial non-Gaussian potential of $\fNL^{\Phi}$ of order
25 would have given (also for an equilateral configuration). As
shown in the text, this number can easily be recovered from
back-of-the-envelop calculations. The first lesson that can be
drawn from this result is that it gives a signal larger than what
a model with a primordial $\fNL$ of order unity would give! The
second lesson is that the $l$-dependence of the bispectrum is
expected to be different from the one induced by primordial mode
couplings. It is expected to have a specific shape as encoded in
the CDM kernel expression.\\

In conclusion, this work offers a breakthrough insight into the
physics of CMB in the non-linear regime and on small angular
scales. It identifies what is, as we argued, the main small scale
contribution of the bispectrum, hence filling the gap with the
standard results that have been obtained in the weakly non-linear
regime of gravitational clustering of dark matter. We did not
check this result against a  (yet non-existing) full second order
Boltzmann code, and this is probably desirable, but we argue that,
given the amplitude of the effects, all other contributions will
be subdominant.  With such a large signal, detection of this
bispectrum should be easily within reach of future CMB experiments!\\

{\bf Acknowledgements:} We thank J. Martin-Garcia for his help in
using the tensorial perturbation calculus package xPert~\cite{xact} that was
used to derive the second order expressions of this paper. We also thank
G. Faye, Y. Mellier, S. Prunet, D. Spergel and N. Aghanim for many discussions.

\bibliography{FNL}
\pagebreak
\appendix
\section{Description of radiation}\label{AppRad}

To describe the evolution of radiation, we use the first
moments of the Boltzmann hierarchy (see e.g.
Refs.~\cite{Hu:1997hp}) including polarisation. The hierarchy
reads
\begin{eqnarray}
\Theta_\ell'&=&k\left[\frac{\ell}{2 \ell -1}\Theta_{\ell-1}-\frac{\ell+1}{2
    \ell +3}\Theta_{\ell+1} \right]\nonumber\\
&&-\tau'\left[\Theta_\ell-\delta_{\ell
    2}\frac{1}{10}\left(\Theta_2-\sqrt{6}E_2 \right)\right]\\
E_{\ell}'&=&k\left[\frac{\sqrt{\ell^2-4}}{2 \ell -1}E_{\ell-1}-\frac{\sqrt{(\ell+1)^2-4}}{2
    \ell +3}E_{\ell+1} \right]\nonumber\\
&&-\tau'\left[E_\ell+\delta_{\ell
    2}\sqrt{6}\left(\Theta_2-\sqrt{6}E_2 \right)\right]
\end{eqnarray}
where the first moments are related to the fluid variables by
$$
 \Theta_0 = \frac{1}{4}\delta_\rad\ ,\quad
 \Theta_1 = -k v_\rad\ , \quad
 \Theta_2 =\frac{5}{12}k^2\pi_\rad\ .
$$
The Boltzmann hierarchy is infinite and we truncated it after the multipole
$\ell=8$, when computing the first order and after $\ell=3$ when computing the
second order. In order to cut the hierarchy without numeric
reflection~\cite{Ma:1995ey}, we use the
free-streaming solution of this hierarchy, and use it to express in the last equation the multipole $\ell+1$ in
function of the multipoles for $\ell$ and $\ell-1$. Explicitly, the closure relation reads
\begin{eqnarray}
\Theta_{\ell+1}&=&\frac{2 \ell+3}{k \eta}\Theta_\ell-\frac{2 \ell+3}{2
  \ell-1}\Theta_{\ell-1}\\
E_{\ell+1}&=&\frac{(2 \ell+3)\sqrt{\ell+1}}{k \eta\sqrt{\ell-1}}E_\ell-\frac{(2 \ell+3)\sqrt{(\ell+3)(\ell+2)}}{(2 \ell-1)\sqrt{(\ell-1)(\ell-2)}}E_{\ell-1}\,.\nonumber
\end{eqnarray}

\section{The Integrated Sachs-Wolfe effect
contribution to the small scale bispectrum}\label{appB}

At linear order, the contribution of the integrated Sachs-Wolfe
effect on small scales is usually small because the time
dependence of the potential vanishes in the matter dominated era.
This is no more the case at second order. It is thus legitimate to
investigate the impact of the time dependence of the second order
gravitational potential on the amplitude of the bispectrum.

At linear order the expression of the temperature anisotropies is
\begin{eqnarray}
 \Theta^{(1)}(\gr{l})=\frac{\sqrt{2 \pi}}{D_\LSS^2}\int\hat g(\gr{k})\,\dd k_r
\end{eqnarray}
while at second order an extra source term should be included. It
is formally given by
\begin{eqnarray}
 \Theta^{(2)}_{\rm ISW}(\gr{l})=\int_{0}^{\etas}
\dd\eta \frac{\dd}{\dd
 \eta}\left\lbrace\Psi^{(2)}[\vx(\eta),\eta]+\Phi^{(2)}[\vx(\eta),\eta]\right\rbrace\,\,\,\,
\label{theta2ISW}
\end{eqnarray}
assuming instantaneous recombination at $\eta=\etas$. To estimate
the magnitude of this effect, we assume that
$\Psi^{(2)}=\Phi^{(2)}$ and that the time dependence is the one
otained in Eq.~(\ref{Phi2MD}),
\begin{equation}
\Psi^{(2)}(\eta)=\frac{\eta^2}{\etas^2}
K_{NL}(\vk_{1},\vk_{2},\etas)\ , \label{KNL}
\end{equation}
where the time dependence has been explicited. Obviously,
expression~(\ref{theta2ISW}) gives an extra term contributing to
the bispectrum. Consistently with our former analysis, let us
evaluate this contribution in the small angle approximation. This
leads to
\begin{eqnarray}
 b_{\vl_{1}\vl_{2}\vl_{3}}^{\rm ISW}&=&
4 \int_{-\infty}^{\infty}\dd k_{r_{1}}\dd k_{r_{2}}\,P(k_{1})\,P(k_{2})\,
K_{NL}(\vk_{1},\vk_{2},\etas)\nonumber\\
&&\hspace{-1cm}\times
\int_{0}^{\etas}\dd\eta_{1}v(\eta_{1})\,\dd\eta_{2}v(\eta_{2})\,\frac{\dd\eta\ \eta}{\etas^2}
g(\vk_{1},\eta_{1})
g(\vk_{2},\eta_{2})\nonumber\\
&&\hspace{-1cm}\times
\exp\left[\ii k_{r_{1}} \etas+\ii k_{r_{2}} \etas\right]+\sym
\label{ISW2}
\end{eqnarray}
that has to be compared with ~Eq.(\ref{blllExp}). 
The integral over $\eta$ can then be performed to give
\begin{eqnarray}
b_{\vl_{1}\vl_{2}\vl_{3}}^{\rm ISW}&\approx&
4 \int_{-\infty}^{\infty}\dd k_{r_{1}}\dd k_{r_{2}}\,P(k_{1})\,P(k_{2})\,
K_{NL}(\vk_{1},\vk_{2},\etas)\nonumber\\
&&\hspace{-1cm}\times
\int_{0}^{\etas}\dd\eta_{1}v(\eta_{1})\,\dd\eta_{2}v(\eta_{2})\,
g(\vk_{1},\eta_{1})
g(\vk_{2},\eta_{2})\nonumber\\
&&\hspace{-1cm}\times w(k_{r_{1}}+k_{r_{2}},\etas)+\sym
\label{ISW2b}
\end{eqnarray}
with
\begin{equation}
w(k,\etas)=\frac{1-\ii k\etas-\exp(\ii k \etas)}{k^2\,\etas^2}\ .
\label{wdef}
\end{equation}
If one examines the UV convergence properties of this expression
(for the integrals over $k_{r}$), it appears that the integral
over $k_{r_{1}}+k_{r_{2}}$ converges at a scale given by the
inverse of $\etas$, e.g.
\begin{equation}
\int_{-\infty}^{\infty}\dd k\ w(k,\etas)=\frac{\pi}{\etas}\ ,
\label{wintegrate}
\end{equation}
due to the oscillatory behaviour of $w$, whereas the integral over
$k_{r_{1}}-k_{r_{2}}$ converges because of the power spectrum
shape and therefore at a scale which is of the order of
$l_{1}/\etas$ or $l_{2}/\etas$ (whichever is smaller).

Thus if the power spectrum is approximated by a power law,
\begin{equation}
P_{\Phi}(k)\sim k^{n_{s}-4}\ ,
\end{equation}
where the spectral index $n_{s}$ varies a priori from $1$ (at very
large scale) to $-3$ at very small ones -- it is a priori of the
order of say $-2$ at the scales of interest -- then the integral
over $k_{r_{1}}-k_{r_{2}}$ leads to the factor,
\begin{eqnarray}
\int_{-\infty}^{\infty} \dd k\,
P_{\Phi}^2(\sqrt{k_{r}^2+l^2/\etas^2})&=&\nonumber\\
&&\hspace{-3cm}
\frac{l^{2 (n_{s}-3)-1} \sqrt{\pi }
   \etas^{7-2 n_{s}} \Gamma
   \left(\frac{7}{2}-n_{s}\right)}{\Gamma
   (4-n_{s})}
\label{DoublePPhi}
\end{eqnarray}
that is $63 \pi/256 \ (\etas/l)^{11}$ for $n_{s}=-2$.

This is to be compared with the amplitude of the intrinsic effects
we have computed. The latter differs in Eq.~(\ref{blllExp})
because of the absence of filtering function $w$. The amplitude of
the bispectrum is then roughly given by,
\begin{eqnarray}
b_{\vl_{1}\vl_{2}\vl_{3}}^{\rm LSS}&\approx& 2R\,
\int_{-\infty}^{\infty}\dd k_{r_{1}}\dd
k_{r_{2}}\,P_{\Phi}(k_{1})\,P_{\Phi}(k_{2})\nonumber \\
&&\quad\times K_{NL}(\vk_{1},\vk_{2},\etas) \label{LSS2}
\end{eqnarray}
so that its amplitude is dominated by the square of
\begin{eqnarray}
\int_{-\infty}^{\infty}\dd k_{r}
P_{\Phi}(\sqrt{k_{r}^2+l^2/\etas^2})&=&\nonumber\\
&&\hspace{-3cm}
\frac{\etas^{3-n_s} l^{2
   \left(\frac{n_s}{2}-1\right)-1} \sqrt{\pi }
   \Gamma
   \left(\frac{3}{2}-\frac{n_s}{2}\right)}{\Gamma
   \left(2-\frac{n_s}{2}\right)}\ ,
\label{int1D}
\end{eqnarray}
which is equal to $(3\pi/8)^2\ (\etas/l)^{10}$ for $n_{s}=-2$. The
ratio of the two contributions scales then as $1/(R l)$ in favor
of the intrinsic effect.

\section{Check of the numerical integration}\label{appC}

We report in this appendix the results of the first order
numerical integration. We first report in Figs.~\ref{fig3} and~\ref{figO1-2} the
evolution of the perturbed quantities where it can be seen that for
$y>y_\star(k)$ the gravitational potential tends to be determined by the cold
dark matter density perturbation. We also report the angular power
spectrum obtained from the flat sky approximation using  the
expression~(\ref{Cl}). The linear dynamics is then used to
calculate the bispectrum arising from a constant primordial $\fNL$
evolved linearly. It can be checked that the form obtained on
Fig.~\ref{Cleps} is completely consistent with the literature
(see e.g. Ref.~\cite{Komatsu:2002db}).
\begin{widetext}
\begin{figure*}[htb]
\includegraphics[width=7cm]{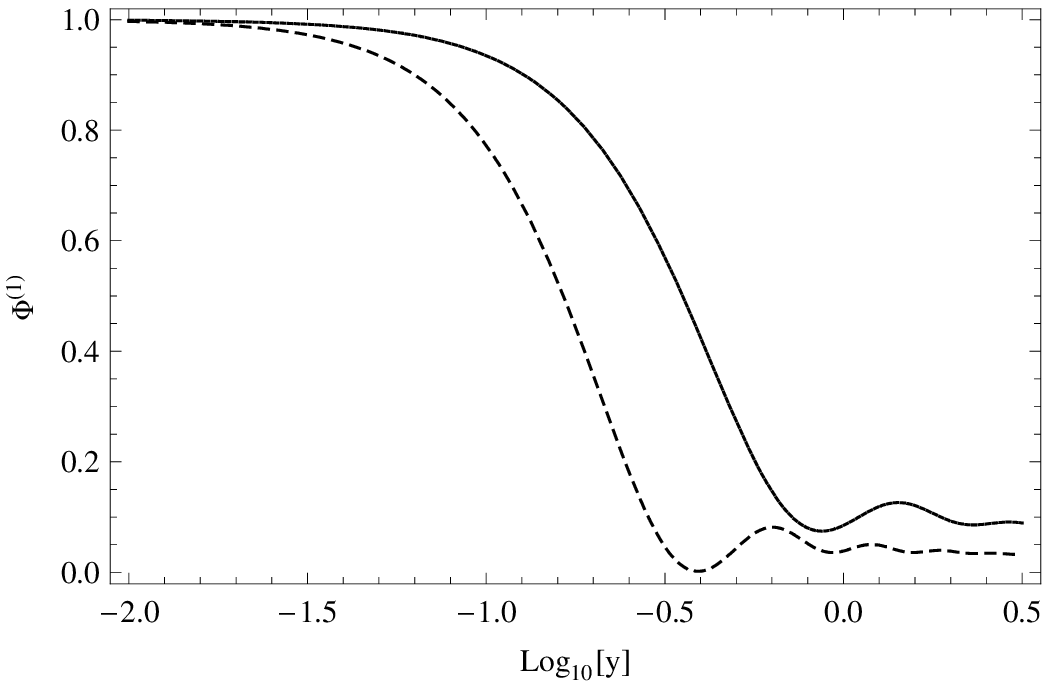}
\includegraphics[width=7cm]{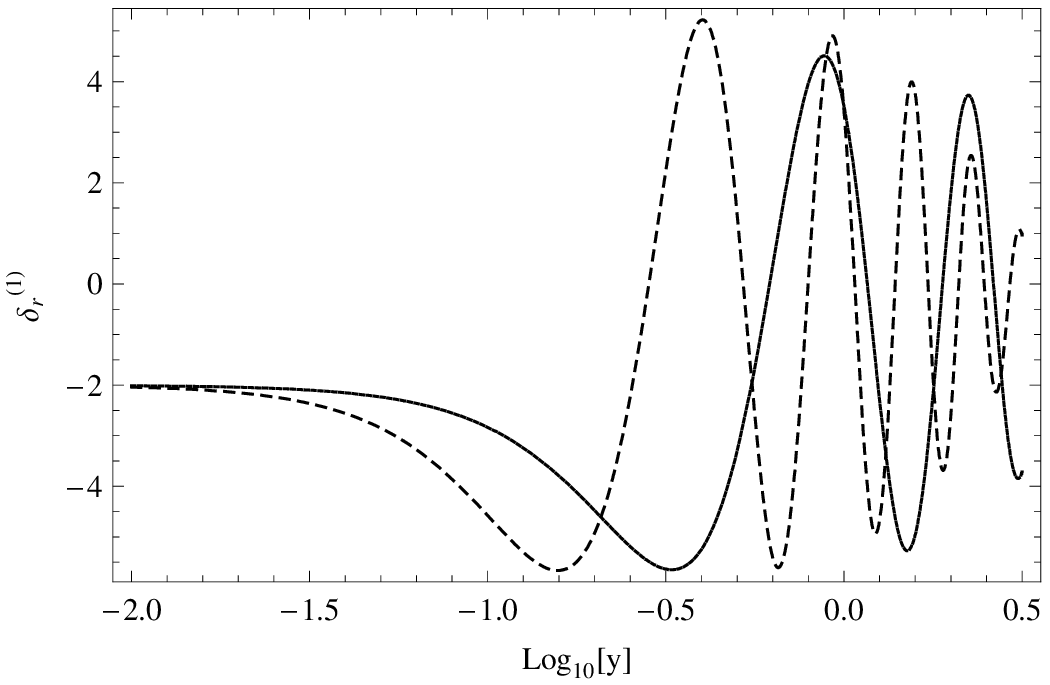}

\includegraphics[width=7cm]{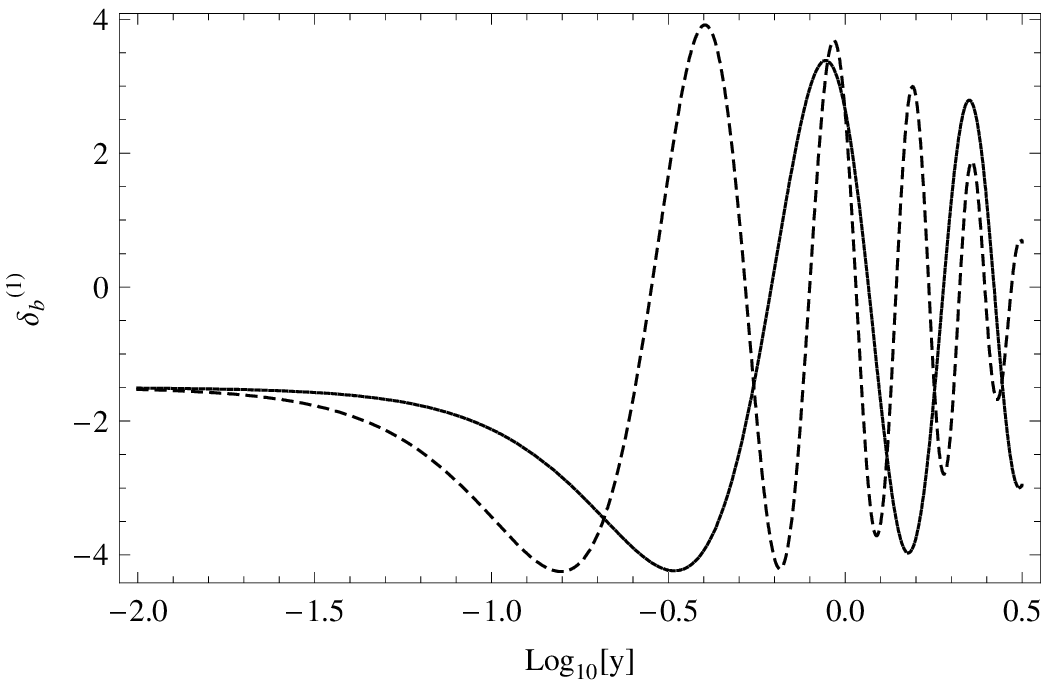}
\includegraphics[width=7cm]{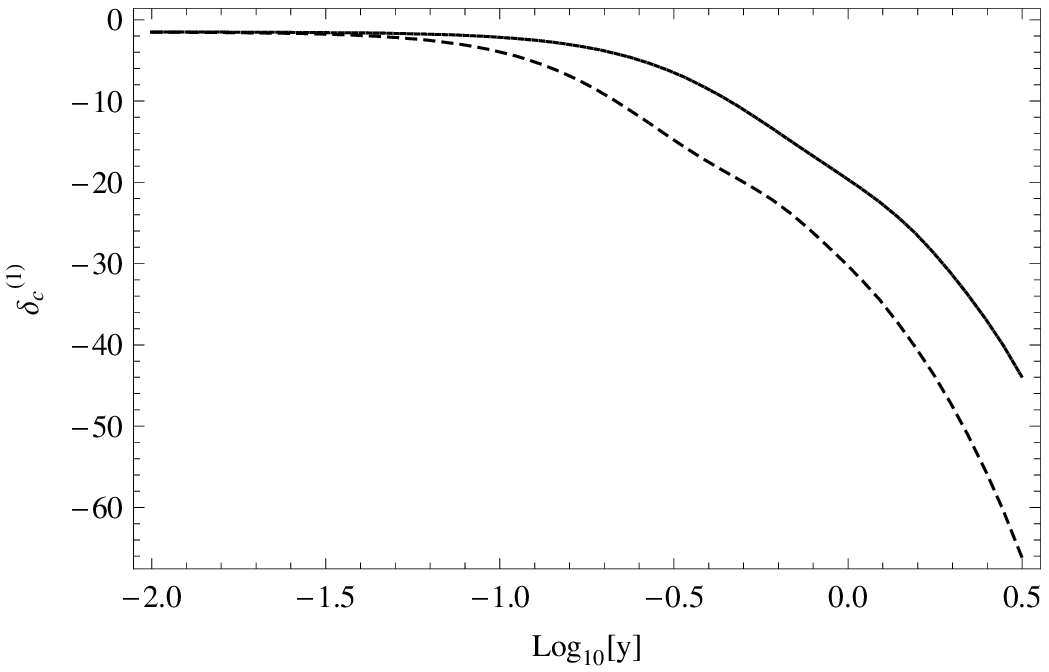}
\caption{From top to bottom, left to right: evolution of the Bardeen
  potential, $\Phi$, and the density contrast $\delta$
  for respectively radiation, baryons and cold dark
  matter. The solid line corresponds to $k= 10k_\EQ$ and
  the dashed line corresponds to $k= 20k_\EQ$.}\label{fig3}
\end{figure*}

\begin{figure}[htb]
\center
\includegraphics[width=8cm]{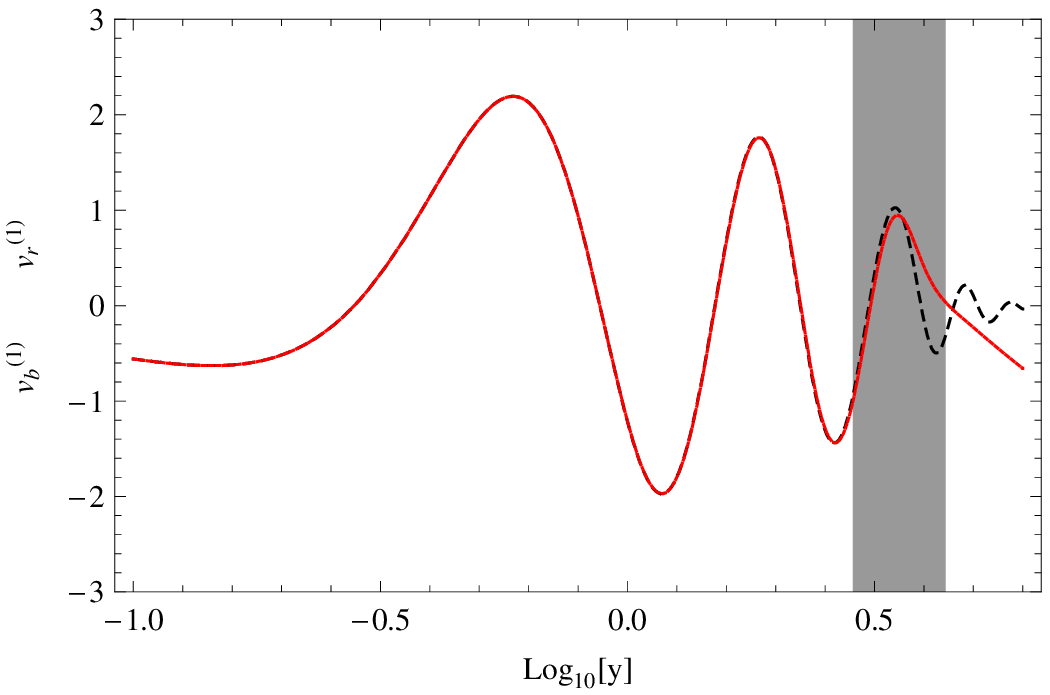}
\includegraphics[width=8cm]{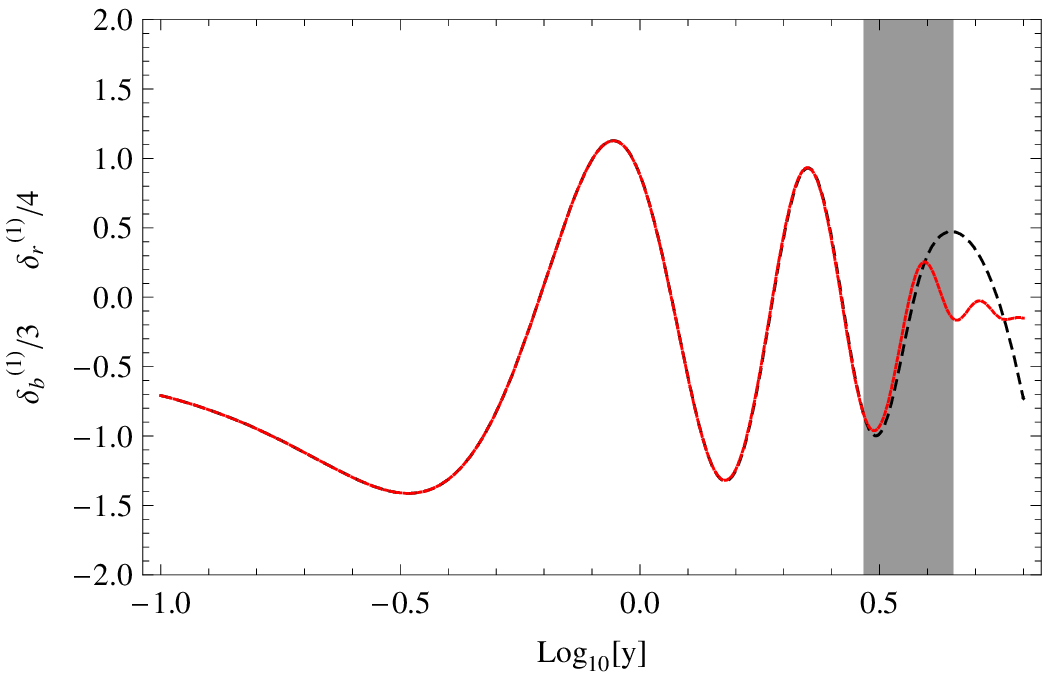}
\caption{Left: Comparison of the baryons and photons velocity
perturbation at first order for $k=10k_\EQ$. It shows that
$v_\rad^{(1)}=v_\baryon^{(1)}$ with a good approximation until
decoupling. Right: Comparison of $\frac14\delta_\rad^{(1)}$ and
$\frac13\delta_\rad^{(1)}$. It can be seen that the adiabaticity
condition holds until recombination, hence justifying the
approximation of \S~\ref{sec3a}.} \label{figO1-2}
\end{figure}

\begin{figure*}[htb]
\includegraphics[width=5cm]{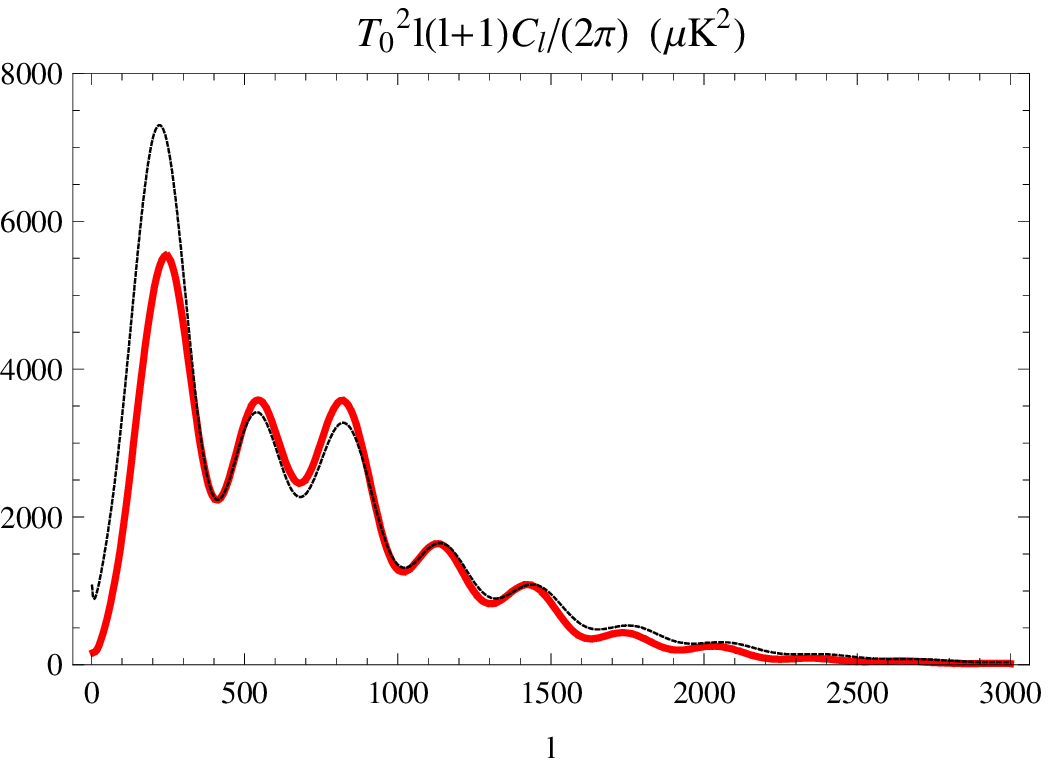}
\includegraphics[width=5cm]{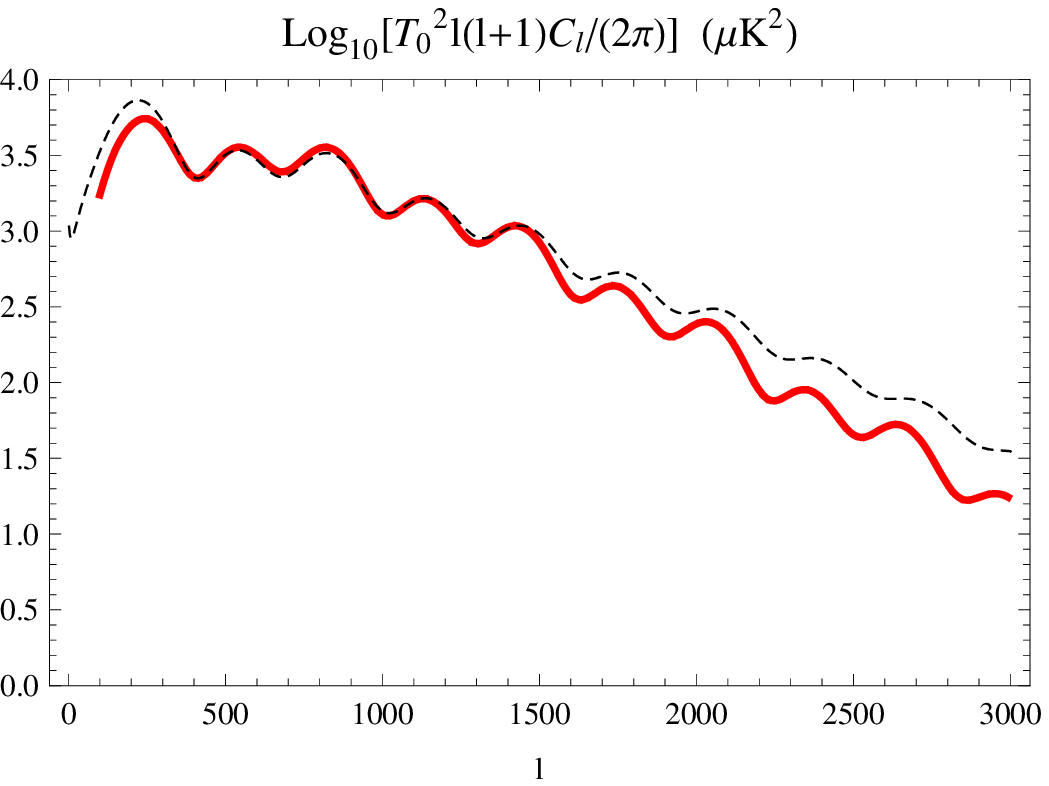}
\includegraphics[width=5cm]{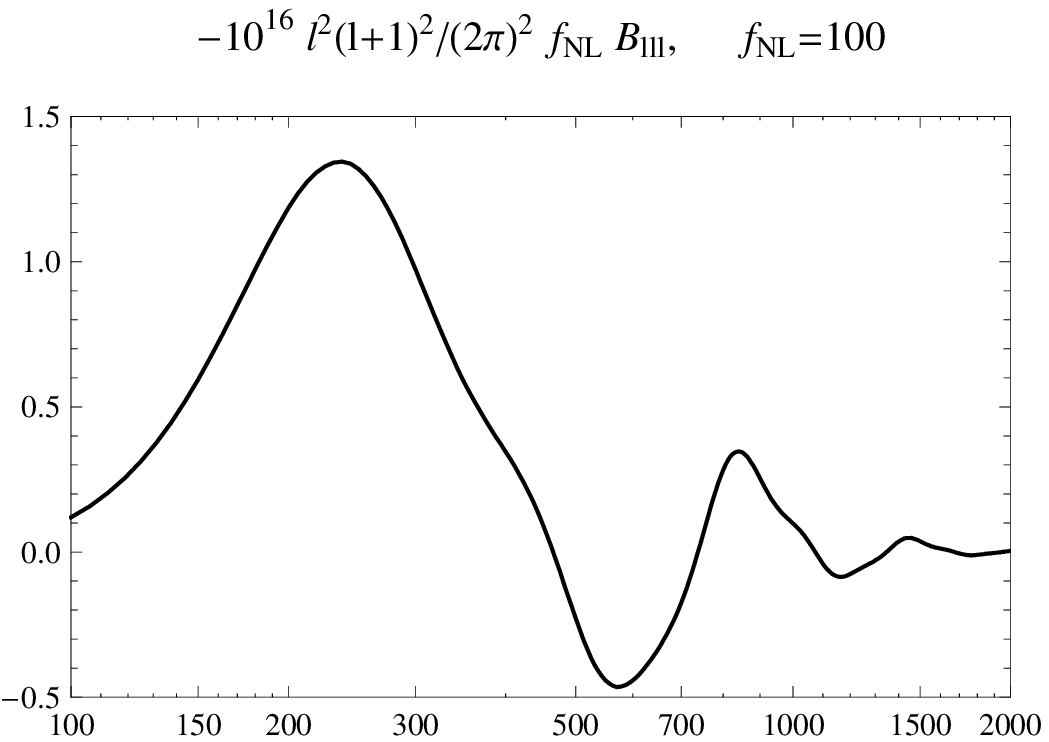}
\caption{Left: in solid red line, the angular power spectrum using
our code and
  the flat sky approximation (which does not includes the late ISW). The black
  dashed line represents the spectrum obtained using
  CAMB, which also takes into account the late ISW, but no reionisation.
  Middle: The precision on small
  scale depends highly on the computation of the visibility function. On small
  scales, our
  transfer function is approximately $30\%$ smaller than the one predicted by
  CAMB and is thus a good approximation.
  Right: the bispectrum obtained from a primordial constant
  $\fNL=100$ evolved linearly.}\label{Cleps}
\end{figure*}
\end{widetext}

\end{document}